\documentclass[useAMS,usenatbib]{mn2e}
\usepackage{aas_macros}
\usepackage{graphicx}
\usepackage{amsmath,amssymb}
\usepackage[T1]{fontenc}
\usepackage{aecompl}
\title[Supernova feedback in a local stratified medium]{Supernova feedback in a local vertically stratified medium: interstellar turbulence and galactic winds}
\author[D. Martizzi et al.]{\parbox[t]{\textwidth}{Davide Martizzi$^{1}$\thanks{E-mail: dav.martizzi@berkeley.edu}, Drummond Fielding$^{1}$,
Claude-Andr\'{e} Faucher-Gigu\`{e}re$^{2}$, Eliot Quataert$^{1}$}\vspace*{6pt}\\
$^{1}$Department of Astronomy and Theoretical Astrophysics Center, University of California, Berkeley, CA 94720-3411, USA.\\
$^{2}$Department of Physics \& Astronomy and Center for Interdisciplinary Exploration and Research in Astrophysics (CIERA),\\ Northwestern University, Northwestern University, Evanston, IL 60208-3112, USA.\\
}

\begin{document}

\maketitle

\label{firstpage}

\begin{abstract}
We use local Cartesian simulations with a vertical gravitational potential to study how supernova (SN) feedback in stratified galactic discs drives turbulence  and launches galactic winds. 
Our analysis includes three disc models with gas surface densities ranging from Milky Way-like galaxies to gas-rich ultra-luminous infrared galaxies (ULIRGs), and two different SN driving schemes (random and correlated with local gas density). 
In order to isolate the physics of SN feedback, we do not include additional feedback processes. 
We find that, in these local box calculations, SN feedback excites relatively low mass-weighted gas turbulent velocity dispersions $\approx3-7$ km s$^{-1}$ and low wind mass loading factors $\eta \lesssim 1$ in all the cases we study.   The low turbulent velocities and wind mass loading factors predicted by our local box calculations are significantly below those suggested by observations of gas-rich and rapidly star-forming galaxies; they are also in tension with global simulations of disc galaxies regulated by stellar feedback. 
Using a combination of numerical tests and analytic arguments, we argue that local Cartesian boxes cannot  predict the properties of galactic winds because they do not capture the correct global geometry and gravitational potential of galaxies. The wind mass loading factors are in fact not well-defined in local simulations because they decline significantly with increasing box height.  More physically realistic calculations (e.g., including a global galactic potential and disc rotation)  will likely be needed to fully understand disc turbulence and galactic outflows, even for the idealized case of feedback by SNe alone.
\end{abstract}

\begin{keywords}
galaxies: general -- galaxies: formation -- galaxies: evolution -- galaxies: ISM -- ISM: supernova remnants -- methods: numerical
\end{keywords}

\section{Introduction}
The evolution of galaxies is strongly influenced by stellar feedback processes, such as ionizing radiation from young massive stars, stellar winds, and supernovae (SNe). 
In many circumstances, SNe are believed to dominate the energy and momentum injection in the interstellar medium \citep[ISM; e.g.,][]{1977ApJ...218..148M, Dekel1986, 1997ApJ...481..703S, 2011ApJ...731...41O, 2013MNRAS.433.1970F}. 
As a result, SNe play a key role in exciting interstellar turbulence \citep[e.g.,][]{2006ApJ...638..797D, 2006ApJ...653.1266J, 2009ApJ...704..137J} and in accelerating galactic winds from star-forming galaxies \citep[e.g.,][]{2005ARA&A..43..769V, 2009ApJ...697.2030S, 2012MNRAS.421.3522H, 2015MNRAS.454.2691M}. 
SNe also strongly influence the dynamics and phase structure of the ISM by inflating bubbles of hot, diffuse gas \citep[e.g.,][]{1977ApJ...218..148M} and by accelerating relativistic cosmic rays \citep[e.g.,][]{1978ApJ...221L..29B, 2016ApJ...816L..19G}. 

Galaxy formation models that do not include strong stellar feedback predict galaxies with star formation rates that are a factor $\sim100$ too high \citep[e.g.,][]{2011MNRAS.417..950H, 2013ApJ...770...25A} relative 
to observations \citep[][]{1998ApJ...498..541K, 2010MNRAS.407.2091G}.  They also form too many stars overall by a factor 
$\sim5$ to $>10^{3}$ \citep[e.g.,][]{1991ApJ...379...52W, 2009MNRAS.396.2332K, 2010ApJ...710..903M, 2010ApJ...717..379B, 2011MNRAS.417.2982F} and fail to explain the observed distribution of heavy elements in 
the intergalactic medium \citep[e.g.,][]{2001ApJ...561..521A, 2006MNRAS.373.1265O, 2010MNRAS.409..132W}.  
While it is clear that stellar feedback is an essential ingredient of galaxy formation models \citep[e.g.,][]{2015ARA&A..53...51S}, even modern cosmological simulations generally do not have sufficient resolution to self-consistently capture stellar feedback. 
To develop realistic subgrid prescriptions to model stellar feedback in cosmological simulations, it is necessary to develop a better understanding of how different feedback processes couple to the scales that are resolved in these simulations. 

Over the past few years, cosmological hydrodynamic simulations have progressed to the point that the state-of-the-art simulations using the zoom-in method \citep[][]{1985PhDT.........7P, 1993ApJ...412..455K} can achieve mass and spatial resolutions better than $\sim 1,000$ M$_{\odot}$ and $\sim10$ pc, respectively, for Milky Way-mass galaxies evolved to the present time (e.g., Guedes et al. 2011; Stinson et al. 2013; Hopkins et al. 2014; Marinacci et al. 2014; Agertz \& Kravtsov 2015; Wetzel et al., in prep.\nocite{2011ApJ...742...76G, 2013MNRAS.428..129S, 2014MNRAS.445..581H, 2014MNRAS.437.1750M, 2015ApJ...804...18A}). 
{ Such simulations directly resolve some of the key processes in the ISM of individual galaxies, such as the formation of giant molecular clouds (GMCs). 
This is significant because the formation of GMCs is likely the rate limiting step for star formation in galaxies \citep[e.g.,][]{2011MNRAS.417..950H, 2013MNRAS.433.1970F}. However, cosmological simulations are not yet capable of directly resolving the formation of individual stars or the full details of how stellar feedback couples to the ISM. 
}
To fully take advantage of the resolution now afforded by cosmological simulations, it is thus critical to develop a subgrid model for SN feedback that accurately captures the effects of SNe on resolved scales.

In \cite{2015MNRAS.450..504M}, we performed high-resolution simulations of isolated supernova remnants (SNRs) in an inhomogeneous ISM and provided fitting formulae for the injection of momentum and thermal energy from individual SNe as a function of ISM density, metallicity, and the size of the SNR.
These fits can be implemented by injecting both residual thermal energy and radial momentum at the resolution scale of cosmological simulations. Using periodic box simulations with randomly seeded SNe, \cite{2015MNRAS.450..504M} verified that this subgrid prescription for SN feedback accurately predicts the thermal energy content and turbulent velocity dispersion of the gas for a given mean gas density and volumetric SN rate. 
A simple variant of this scheme has been used to model SN feedback in the FIRE (Feedback In Realistic Environments) cosmological simulations,  
which have proved successful at reproducing a range of observations for galaxies below $\sim L^{*}$  \citep[e.g.,][]{2014MNRAS.445..581H, 2015MNRAS.449..987F, 2015arXiv150402097M}. 
Several recent studies by other groups performed similar high-resolution simulations of SNRs evolving in an inhomogeneous medium and reached broadly consistent conclusions regarding the momentum boost obtained at the end of the Sedov-Taylor phase \citep{2015ApJ...802...99K, 2015MNRAS.451.2757W, 2015MNRAS.449.1057G, 2015ApJ...814....4L};  subgrid SN schemes analogous to those we developed in \citet{2015MNRAS.450..504M} have also been contemporaneously developed by other groups \citep{2015ApJ...809...69S, 2015MNRAS.451.2900K}. 

Idealized simulations of isolated SNRs -- or simulations of multiple SNe assuming periodic boundary conditions --  have a number of limitations, which introduce significant uncertainties for understanding the impact of SN feedback on galaxies.
In real galaxies, the energy and momentum returned by SNe contribute both to ISM turbulence and to driving galactic winds.   Simulations with periodic boundary conditions clearly cannot address how efficiently winds are driven. 
In this paper, we thus extend the idealized inhomogeneous SN feedback study of \cite{2015MNRAS.450..504M} to include the effects of vertical stratification in galactic discs by imposing a background gravitational potential as well as outflow boundary conditions in the vertical direction. 
We continue to adopt the local box approximation, focusing on $\sim$ 1 kpc$^{3}$ patches of galactic discs.   

Several previous SN feedback studies have included vertical stratification in local boxes \citep[][]{2006ApJ...638..797D, 2006ApJ...653.1266J, 2009ApJ...704..137J, 2013MNRAS.429.1922C, 2013ApJ...776....1K, 2015ApJ...815...67K, 2016MNRAS.456.3432G}. 
Our analysis complements various subsets of these previous studies in different ways. Our simulations implement the subgrid SN model developed in \cite{2015MNRAS.450..504M}, allowing us to compare in a well-controlled manner our new results with those obtained before in periodic boxes.   We also systematically compare models with three different gas surface densities as a proxy for galaxies ranging from ones similar to the Milky Way to gas-rich ultra-luminous infrared galaxies (ULIRGs); and for each model, we study two different SN driving schemes (random and correlated with local gas density). {\cite{2016MNRAS.456.3432G} recently published a closely related study, which explored a larger set of SN seeding models. \cite{2016MNRAS.456.3432G} included gas self-gravity and magnetic fields, which we neglect, but focused on a Milky Way-like disc.} The biggest difference between this paper and previous work lies primarily, however, in how we design the simulation catalog and analyse the results.    First, we quantitatively assess what fraction of the SN feedback energy goes into winds vs. turbulence. We also discuss the implications for analytic models of the Kennicutt-Schmidt (K-S) law based on a balance between SN feedback and gravity \citep[][]{2011ApJ...731...41O, 2013MNRAS.433.1970F}.

Finally, we analyse the sensitivity of our results to vertical boundary conditions and identify limitations generic to local Cartesian box calculations, which appear to have been under-appreciated in the context of SN feedback studies. 
We argue that local Cartesian boxes (both ours and other ones in the literature) are inherently limited in their ability to quantitatively predict galactic wind properties.  Evidence for this includes (i) the strong dependence of  wind properties on box height and (ii) an analytic derivation showing that steady adiabatic SNe-heated winds do not undergo a standard subsonic to supersonic transition in plane-parallel geometry (Appendix \ref{appendix:C}).  These limitations of local Cartesian box calculations suggest that further increasing the physical realism of local calculations or moving to global galaxy models will be needed to properly understand how SN feedback generates galactic winds.   In addition, our results strongly suggest that fits to wind properties derived from local simulations \citep[e.g.,][]{2013MNRAS.429.1922C} could produce misleading results if used as a basis for modeling SN feedback in cosmological simulations or semi-analytic models. 

This paper is structured as follows. \S \ref{sec:setup} describes our simulation methodology. \S 3 discusses the global dynamics and phase structure of the ISM in the simulations. \S 4 summarises our main results concerning the energetics of SN feedback. In \S 5, we focus on turbulence in the disc and its relationship to analytic K-S law models. \S 6 discusses the effects of the box geometry and boundary conditions on our results. We conclude in \S 7. 
Appendices contain more details regarding the convergence with resolution and box height of our simulations, as well as an analytic treatment of steady winds in plane-parallel geometry.    

\section{Simulation set-up and methodology}
\label{sec:setup}
Our main simulations use the {\sc ramses} code \citep{Teyssier:2002p451}, an adaptive mesh refinement (AMR) code based on a second-order unsplit Godunov solver \citep{Toro:1994p1151} (in Appendix \ref{appendix:B}, we compare our {\sc ramses} results to tall boxes run with {\sc athena}). We evolve an ideal hydrodynamic fluid in 3D without self-gravity and we adopt the HLLC Riemann solver. 
We initialize the metal mass fraction of the gas to a fixed value $Z=Z_{\odot}=0.02$ {(the hydrogen and helium mass fractions are $X=0.76$ and $Y=1-X-Z$, respectively)}. 
Metal line cooling at $T<10^4$ K and photo-electric heating from dust grains are neglected. 
The box side length of our cubical {\sc ramses} boxes is denoted by $L$. 
We use uniform grids (no mesh refinement) throughout.

\subsection{Vertically stratified disc models}
\label{sec:disc_models}
We model local patches of galactic discs by imposing a fixed vertical  gravitational potential intended to approximate the effects of a stellar disc and a dark matter halo; the self-gravity of the gas is neglected. The specified gravitational potential depends only on the coordinate $z$ normal to the disc plane. The disc mid-plane is located at $z=0$. 
The specific form of the gravitational potential we adopt is the same as in \citet{1989MNRAS.239..605K},
\begin{equation}\label{eq:potential}
 \phi (z)=a_1\left[ \sqrt{z^2+z_{0}^2}-z_0\right]+\frac{a_2}{2}z^2,
\end{equation}
where $a_1$ and $a_2$ are coefficients associated with the disc (stellar) and spheroidal (halo) components, respectively, and $z_0$ is the scale height of the stellar disc. This potential corresponds to an infinite thin disc of stellar surface density $\Sigma_{*}$ embedded in a halo of constant density $\rho_{\rm h}$, such that
\begin{equation}
a_1=2\pi G \Sigma_{*}
\end{equation}
and 
\begin{equation}
a_2= \frac{4\pi G}{3}\rho_{\rm h}.
\end{equation}
We assume that the gas is initially isothermal with temperature $T_0$ and in hydrostatic equilibrium with the prescribed potential. 
We set the initial temperature to $T_0=1.32\times 10^4$ K. Since the initial profile of the gas declines rapidly with increasing $z$, we impose a floor $\rho_{\rm gas,min} = \rho_{\rm gas}(z=250 \hbox{pc})$ on the initial gas density distribution. 
We define an effective gaseous disc scale height $z_{\rm eff}$ such that 
\begin{equation}\label{eq:zeff}
 \Sigma_{\rm gas}=\int_{-L/2}^{+L/2}dz \rho_{\rm gas}(z)=2z_{\rm eff}\rho_0,
\end{equation}
where the $\rho_0$ is the initial mid-plane gas density. 
At later times the disc is supported in part by turbulent pressure, rather than thermal pressure.  As a result, the true effective scale height can differ from the one defined using the initial conditions in equation \ref{eq:zeff}. 
We nevertheless use $z_{\rm eff}$ to define a characteristic height used to measure various quantities for each of our disc models; the true steady-state scale heights are somewhat smaller.
Finally, for each disc model we define the gas mass fraction as
\begin{equation}
\label{eq:gas_fraction}
f_{\rm gas} = \frac{\Sigma_{\rm gas}}{\Sigma_{*}}.
\end{equation}

We simulate three disc models, corresponding to different galaxy types. The parameters for each disc model are given in Table~\ref{tab:parameters1}. The Milky Way-like MW model has $\Sigma_{\rm gas} = 5$ M$_{\odot}/$pc$^2$, $f_{\rm gas}=0.088$, and $z_0=180$ pc. {These parameters are similar to those summarized by \cite{2015ApJ...814...13M} for the solar neighborhood.} The ULTRA-MW model, with $\Sigma_{\rm gas} = 50$ M$_{\odot}/$pc$^2$, is our reference configuration. To achieve a similar gas fraction as in the MW model, the parameters $a_1$ and $a_2$ are increased by a factor $\approx10$, while $z_0$ is kept fixed at $180$ pc. 
{The ULTRA-MW model is representative of a galaxy whose gas surface density \emph{and} total mass are larger than the Milky Way.} 
Finally, we investigate a gas-rich ULIRG model with gas surface density $\Sigma_{\rm gas} = 500$ M$_{\odot}/$pc$^2$ representative of ultra-luminous infrared galaxies (though not as extreme as, e.g., the nucleus of Arp 220). The scale height of the gravitational potential in the ULIRG model, $z_0=300$ pc, is higher than for the MW and ULTRA-MW models, as is its gas fraction $f_{\rm gas}=0.496$. However, the ULIRG model has an initial hydrostatic gaseous disc slightly thinner than the ULTRA-MW model.

\begin{table*}
\centering
{\bfseries Initial parameters for our vertically stratified disc models}
\makebox[\linewidth]{\scriptsize
\begin{tabular}{lcccccccccc}
\hline
Disc model & $\Sigma_{\rm gas}$ [M$_{\odot}/$pc$^2$] & $f_{\rm gas}$ & $z_{\rm eff}$ [pc] & $a_1$ [kpc Myr$^{-2}$] & $a_2$ [Myr$^{-2}$] & $z_0$ [pc] & $\sqrt{2 \phi(z=\hbox{500 pc})}$ [km/s] \\
\hline
 MW & 5 & 0.088 & 165 & $1.42\times 10^{-3}$ & $5.49\times 10^{-4}$ & 180 & 33 \\
 ULTRA-MW & 50 & 0.100 & 99 & $1.26\times 10^{-2}$ & $4.87\times 10^{-3}$ & 180 & 100 \\
 ULIRG & 500 & 0.496 & 87 & $2.78\times 10^{-2}$ & $9.80\times 10^{-3}$ & 300 & 130 \\ 
\hline
\end{tabular}
}
\caption{ Symbols are defined in \S \ref{sec:disc_models}. The last column indicates the  escape speed needed to reach the box edge ($|z| =$ 500 pc) from the mid-plane in our fiducial simulations.   All discs are initialized in hydrostatic equilibrium at an initial temperature $T_{0}=1.32\times 10^4$ K.   }\label{tab:parameters1}
\end{table*}

Our simulations use periodic boundary conditions on the sides (along the $x$ and $y$ axes) and outflow boundary conditions (gradients of density, velocity, pressure, metallicity, and gravitational potential set to zero) at the top and bottom of the domain. Mass flow into the box from the ghost zones at the $\pm z$ boundaries is not allowed ($v_{z}$ is prevented from becoming negative at the top boundary, or positive at the bottom boundary). In our fiducial simulations, we do not explicitly enforce hydrostatic balance at the top and bottom of the domain; we verified that this choice does not influence our results. To assess the convergence of our results with spatial resolution, we run simulations at three different resolutions. 
The lowest resolution, labeled `-L7' uses a Cartesian mesh with $(2^7)^3=128^3$ cells (cell size $\Delta x=7.8$ pc). 
The fiducial resolution, labeled `-L8' uses a Cartesian mesh with $(2^8)=256^3$ cells ($\Delta x=3.9$ pc).
The highest resolution, labeled `-L9' uses a Cartesian mesh with $(2^9)=512^3$ cells ($\Delta x=1.95$ pc). 
Appendix \ref{appendix:A} summarises convergence tests and shows that for the global quantities we focus on in this paper, our results are well converged with respect to spatial resolution at the fiducial `-L8' resolution.

Our fiducial simulations use a box side length $L=1$ kpc. To test the robustness of our results with respect to box size, we also ran cubical {\sc ramses} simulations with $L=2$ kpc (labeled `-LB' ). In addition, we ran a ``tall box'' simulation of height 5 kpc with {\sc athena}, described in Appendix~\ref{appendix:B}. 
Overall, we find that the disc properties predicted by our simulations are robust to box size but that the detailed wind structure does depend on box size.  In particular, the wind mass outflow rate decreases with increasing box height so there is no well defined mass outflow rate in local Cartesian simulations of SNe-driven galactic winds (see \S \ref{sec:bc_dep}).

\subsection{Supernova seeding schemes}
\label{sec:SN_seeding}
To model SNe, we inject energy and momentum in spheres of fixed radius $R_{\rm inj}=2\times \Delta x$ for each SN following the subgrid model developed by \cite{2015MNRAS.450..504M} based on high-resolution simulations of isolated SNRs. Specifically, we use the fits in equation (11) of \cite{2015MNRAS.450..504M} as a function of ambient density $n_{\rm H}$ and metallicity $Z$ (evaluated as the means within $R_{\rm inj}$) to determine the radial momentum and residual thermal energy $P_{\rm r}(R_{\rm inj},n_{\rm H}, Z)$ and $E_{\rm th}(R_{\rm inj},n_{\rm H}, Z)$ at the injection radius. 
In the limit in which the cooling radius of SNRs is well resolved by our local disc simulations, this subgrid model reduces to simply injecting $10^{51}$ erg of thermal energy per SN and letting the simulation code explicitly capture the conversion into radial momentum. 
When the cooling radius is not well resolved {(i.e., when it is smaller than or comparable to the injection radius),} the subgrid prescription approximates the thermal energy and momentum deposition based on the isolated SNR fits. 
The radial momentum is injected in a way that enforces the net linear momentum vector deposited to be zero (i.e., radial momentum canceling along opposite directions). 
We also return 3 M$_{\odot}$ of mass into the ambient medium for each SN explosion (a negligible effect on the total mass in the box).

We do not explicitly model star formation but instead explore two different schemes for seeding SNe.
 
In the first scheme (`FX-' suffix for ``fixed''), SNe are randomly seeded in space and time within $|z|  \leq z_{\rm eff}$ (defined in eq.~\ref{eq:zeff}). 
The SN rate is set based on an observationally-inferred Kennicutt-Schmidt law $\dot{\Sigma}_{*} \propto \Sigma_{\rm gas}^{1.4}$ \citep{2007ApJ...671..333K}, where $\dot{\Sigma}_{\rm *}$ is the star formation rate surface density. 
$\dot{\Sigma}_{*}$ is converted to a volumetric SN rate via $\dot{n}_{\rm SN}=\dot{\Sigma}_{*}/(2 z_{\rm eff} m_*)$, where $m_*=100$ M$_{\odot}$ (i.e., one SN per 100 M$_{\odot}$ of newly formed stars). 
Since $z_{\rm eff}$ is typically larger than the steady-state gaseous disc scale height, a relatively large fraction of the SNe explode in regions of relatively low density above or below the disc. 
{The fractions of SNe exploding at $\rho<\rho_0/3$ are $\sim10$\%, $\sim60$\% and $\sim 60$\% for ÄFX-MW, FX-ULTRA-MW and  FX-ULIRG, respectively. 
Such fractions could occur if a high fraction of SNe are produced by runaway OB stars \citep[e.g.,][]{2012ApJ...755..123C}. 
Furthermore, radiative feedback from star formation can clear molecular clouds of their gas before all core collapse SNe have had time to explode \citep[e.g.,][]{2012MNRAS.421.3488H, 2013MNRAS.436.3727D}. 
We stress, however, that our FX model is primarily intended to test the random limit for SN feedback, rather than being motivated in detail by a particular physical model. 
}

In the second scheme (`SC-' suffix for ``self-consistent''), we assume that star formation occurs on a time scale $t_{\rm *}$ that is a multiple of the local dynamical time, $t_{\rm *}=f_{*} t_{\rm dyn}$. We adopt the value $f_{*}=100$, corresponding to a star formation efficiency of one percent per dynamical time. The dynamical time is calculated using the mass density corresponding to the prescribed potential,
\begin{equation}\label{eq:tff}
t_{\rm dyn}(z)=\sqrt{\frac{3\pi}{16G\rho(z)}}, 
\end{equation}
where
\begin{equation}
\rho(z)=\frac{1}{4\pi G}\nabla^2\phi(z)
\end{equation}
(in the disc, the stellar component dominates the potential). 
For any given cell, the volumetric SN rate is then given by
\begin{equation}\label{eq:nsnloc}
\dot{n}_{\rm SN,loc}=\frac{\rho_{\rm gas}}{m_* t_*}
\end{equation}  
As for the FX scheme, SNe are seeded randomly but with a probability per time step proportional to the local gas density. 
We do not allow more than one SN to explode in a single cell per time step. 
In the SC scheme, SNe preferentially explode where the gas density is high, such as near the mid-plane and around density peaks. We do not, however, enforce an explicit cut on the distance from the mid-plane at which SNe are allowed to occur. 
Note also that this scheme is not as extreme as the `peak driving' scheme of \citet{2016MNRAS.456.3432G}, in which SNe are seeded \emph{only} at the highest density peaks. 

\begin{table*}
\centering
{\bfseries Parameters of the local disc simulations}
\makebox[\linewidth]{\scriptsize
\begin{tabular}{lcccccc}
\hline
Simulation name & Disc model & $L$ [kpc] & Number of cells & $\Delta x$ [pc] & SN scheme & $\langle\dot{n}_{\rm SN}\rangle$ [Myr$^{-1}$ kpc$^{-3}]$ \\ 
\hline
FX-MW-L8 & MW & 1 & $256^3$ & 3.9 & Fixed for $|z|\le z_{\rm eff}$ & 30 \\
FX-ULTRA-MW-L7 & ULTRA-MW & 1 & $128^3$ & 7.8 & Fixed for $|z|\le z_{\rm eff}$ & 1,500 \\
FX-ULTRA-MW-L8 & ULTRA-MW & 1 & $256^3$ & 3.9 & Fixed for $|z|\le z_{\rm eff}$ & 1,500 \\
FX-ULTRA-MW-L9-LB & ULTRA-MW & 2 & $512^3$ & 3.9 & Fixed for $|z|\le z_{\rm eff}$ & 1,500 \\
FX-ULTRA-MW-L9 & ULTRA-MW & 1 & $512^3$ & 1.95 & Fixed for $|z|\le z_{\rm eff}$ & 1,500 \\
FX-ULIRG-L8 & ULIRG & 1 & $256^3$ & 3.9 & Fixed for $|z|\le z_{\rm eff}$ & 43,000 \\
SC-MW-L8 & MW & 1 & $256^3$ & 3.9 & $\propto \rho_{\rm gas}$ & 70 \\
SC-ULTRA-MW-L8 & ULTRA-MW & 1 & $256^3$ & 3.9 & $\propto \rho_{\rm gas}$ & 1,700 \\
SC-ULTRA-MW-L9 & ULTRA-MW & 1 & $512^3$ & 1.95 & $\propto \rho_{\rm gas}$ & 1,700 \\
SC-ULIRG-L8 & ULIRG & 1 & $256^3$ & 3.9 & $\propto \rho_{\rm gas}$ & 25,000 \\
\hline
\end{tabular}}
\caption{ $L$ is the box side length, $\Delta x$ is the cell size, and $\langle \dot{n}_{\rm SN} \rangle$ is the time averaged SN rate per unit volume in the region $|z| \leq z_{\rm eff}$. }\label{tab:parameters2}
\end{table*}

The parameters of all the {\sc ramses} simulations we analyse in this paper are summarised in Table~\ref{tab:parameters2}.


\section{Global Disc Structure \& Dynamics}
\subsection{Overview and dependence on SN scheme}
All of the simulations  begin with an initial transient phase that lasts $\sim 20$ Myr, comparable to the disc dynamical time.  After $\sim 20$ Myr turbulent 
motions in the bulk of the disc reach a statistical steady state.   In the ULTRA-MW and ULIRG cases, a steady outflow  is  also achieved. The MW model also develops a wind at late times, 
but it is further from steady state (the dynamical time for the MW model is a factor $\sim$ few longer than  the other models). 

\begin{figure*}
	\includegraphics[width=0.99\textwidth]{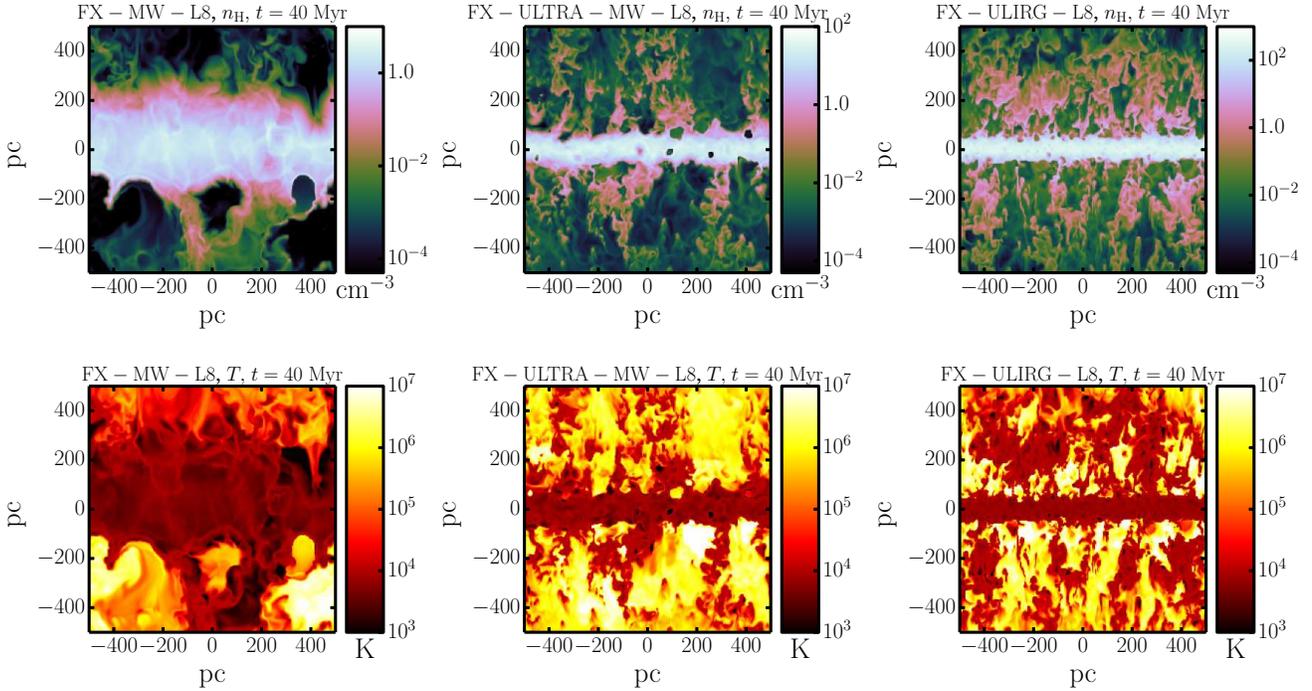}
\caption{Density (top) and temperature (bottom) maps for the fiducial simulations with spatially and temporally random SNe (but a statistically fixed SN rate). The maps are for a slice  
2 cells thick passing through the centre of the simulation box. Left column: 
FX-MW-L8. Central column: FX-ULTRA-MW-L8. Right column: FX-ULIRG-L8.   The corona and wind are noticeably less well-developed in the MW model relative to the two higher surface density (and SN rate) models.   SN remnants are visible in the disc as low density cool bubbles. }\label{fig:maps_fx}
\end{figure*}

\begin{figure}
\begin{center}
    \includegraphics[width=0.49\textwidth]{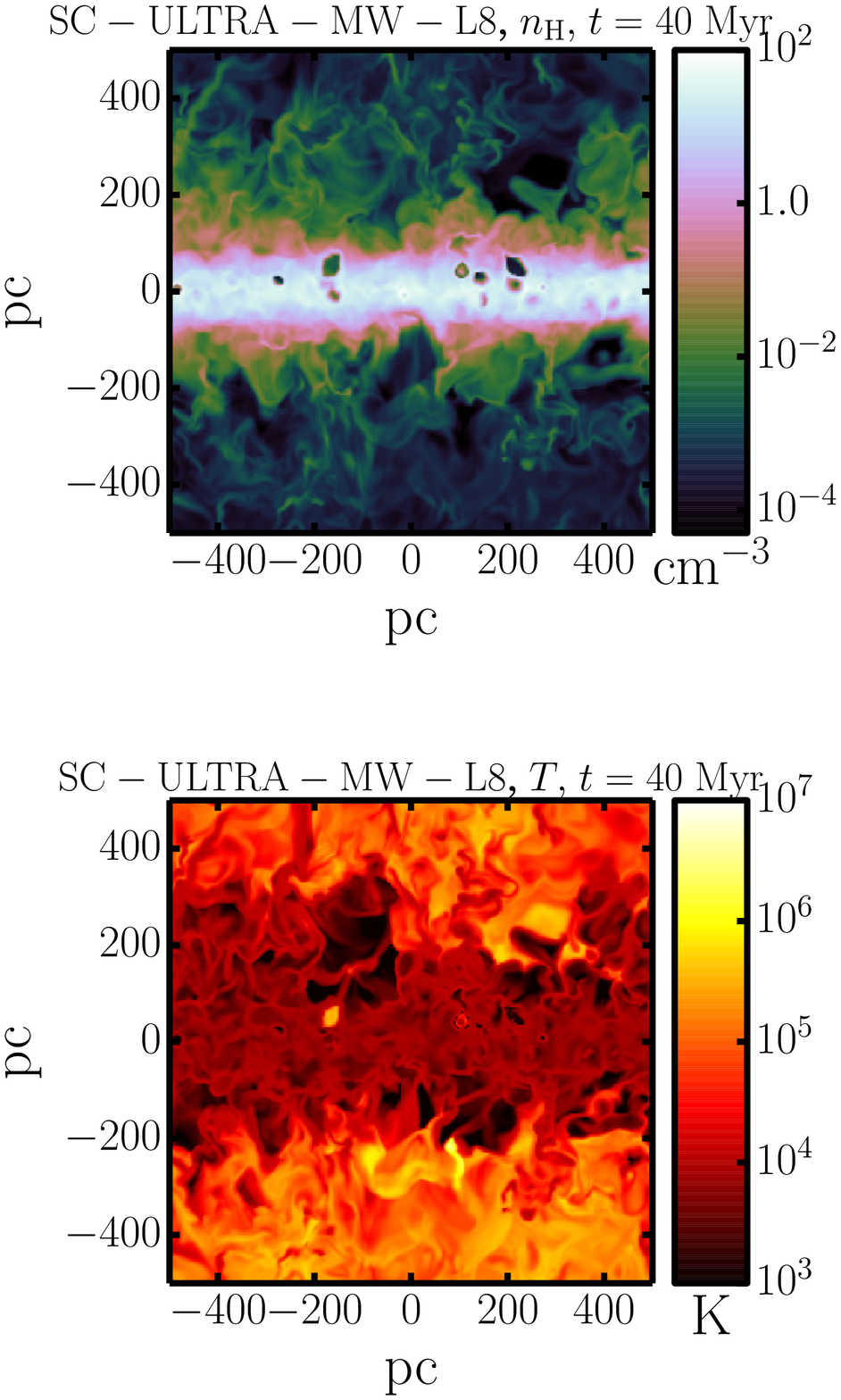}
\end{center}
\caption{ Density (top) and temperature (bottom) maps for the SC-ULTRA-MW-L8 simulation with a SN rate $\propto \rho$. The maps are for a slice 2 cells thick passing through the centre of the simulation box.  The wind and corona are noticeably cooler and less dense than in Figure \ref{fig:maps_fx}.  This is because the SN rate $\propto \rho$ used in the SC simulations leads to more radiative losses than the random driving in Figure \ref{fig:maps_fx}.}\label{fig:maps_sc}
\end{figure}

Figure~\ref{fig:maps_fx} shows  density and temperature maps in the quasi-steady state at  $t = 40$ Myr for the FX SNe simulations 
at our fiducial resolution.   For comparison, Figure \ref{fig:maps_sc} shows the same maps for the ULTRA-MW run with the SC supernova driving model.    The disc structure is characterised by a mid-plane temperature $T\sim 10^4$ K and a hotter `corona/atmosphere.'   This is particularly true for the ULTRA-MW and ULIRG simulations, for which the corona is highly inhomogeneous with large density and temperature fluctuations.  By contrast, in the MW case the hot atmosphere is not as well developed.    The cooler, 
high density filaments visible in the atmosphere in all of the  simulations are typically associated with gas that is cooling and falling back to the disc after being pushed out by the outflow. 

The comparison of Figure~\ref{fig:maps_fx} (FX-runs) and Figure~\ref{fig:maps_sc} (SC-ULTRA-MW-L8) highlights the sensitivity of the corona dynamics to the SN seeding scheme, 
i.e. on the location of the SNe. In the FX case, SNe explode with uniform probability in a horizontal slab of fixed thickness (Table~\ref{tab:parameters2}). The 
probability that a SN explodes in a low density region is relatively high. Since SNe that explode in low density regions are more efficient at heating the gas, 
the FX runs typically have a much hotter and more dynamic atmosphere. In the SC runs, by contrast, SNe preferentially explode in higher density regions close to the mid-plane; as a result,  SNe
heating is less efficient in the SC case and the atmosphere is colder than in the FX case.   This translates into a much weaker outflow (\S \ref{sec:energetics}).   

\begin{figure*}
    \includegraphics[width=0.99\textwidth]{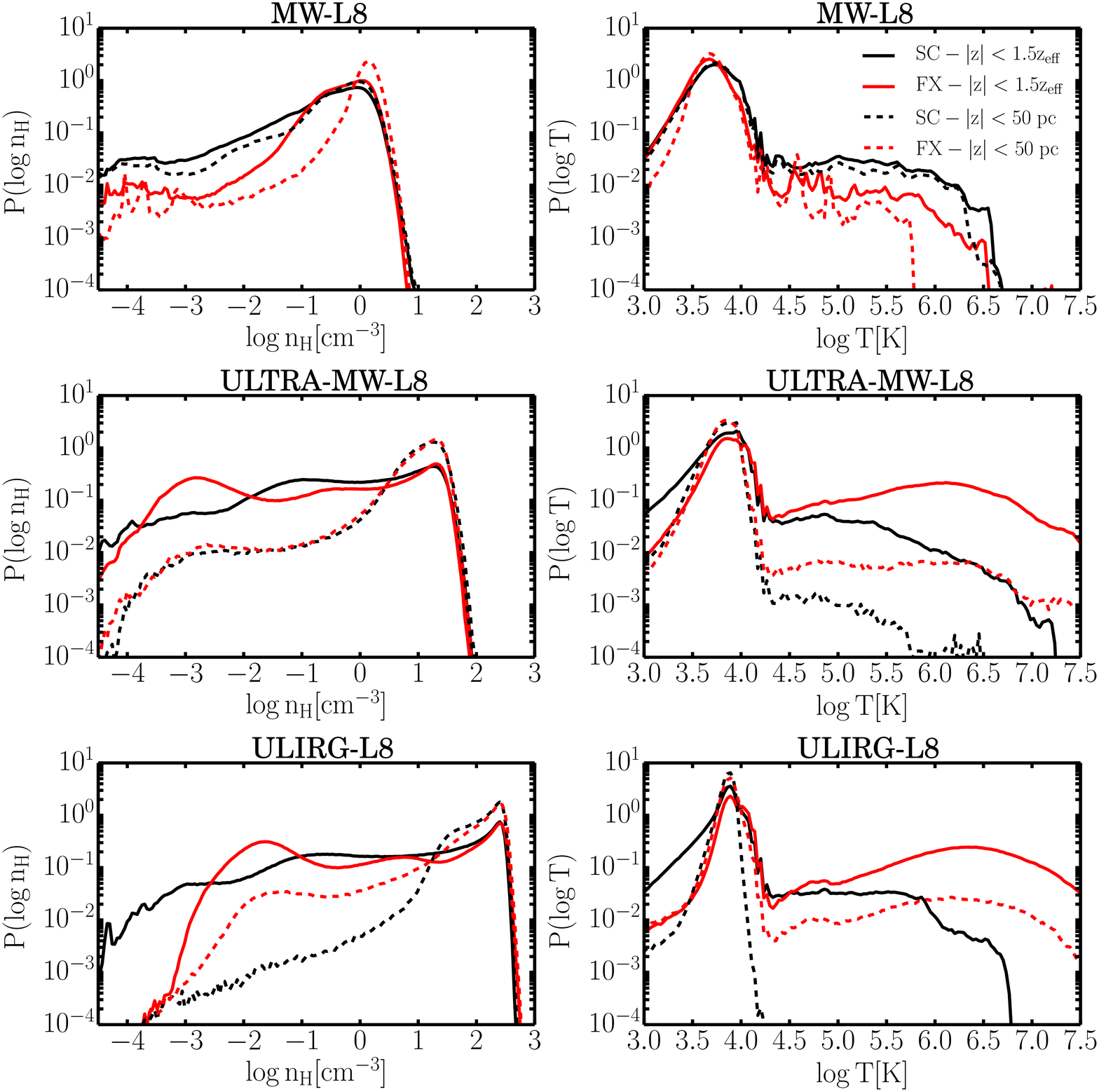}
\caption{Volume weighted probability functions of $\log n_{\rm H}$ (left column) and $\log T$ (right column). The probability functions are measured in the region 
$-1.5 z_{\rm eff}<z < +1.5z_{\rm eff}$ (solid lines) and $-50 \hbox{ pc} < z < 50 \hbox{ pc}$ (dashed lines) and are time averaged over $30 \hbox{ Myr} < t < 40 \hbox{ Myr}$. Each row is a different configuration: MW (top), ULTRA-MW (middle) and 
ULIRG (bottom). SC runs are shown in black, FX runs in red.   In all of the models, the bulk of the disc has a density of order the mean density of the disc and a temperature of $\sim 10^4$ K but there is a significant tail to the PDFs associated with lower density gas in SNRs.   }\label{fig:phase_distr}
\end{figure*}

\subsection{Phase Structure of the Discs}
In the density and temperature maps in Figures~\ref{fig:maps_fx} and \ref{fig:maps_sc} one can identify 
regions where SNe recently exploded close to the disc mid-plane. SNRs can be observed in two different phases: (I) the hot energy conserving Sedov-Taylor phase (II) the  momentum-conserving phase, after the SN remnant has cooled  to low temperatures but has yet to mix into the ambient ISM.\footnote{Adiabatic expansion of remnants can decrease the temperature below the minimum temperature allowed by the cooling curve.}  The hot phase of a SNR is comparatively rare  in these simulations due to the short cooling times (high densities) that characterise the disc models we analyze. It is much more common to find cool, low density cavities characteristic of SNRs that have already reached their momentum conserving phase.   To understand this quantitatively, it is useful to recall that the cooling radius of a SNR at solar metallicity is $R_c \simeq 8 (n/10 \, {\rm cm^{-3}})^{-0.42} \, {\rm pc}$ (e.g., \citealt{2015MNRAS.450..504M}).    Thus unless $n \ll 10 \, {\rm cm^{-3}}$, the SNRs are relatively small in their energy conserving phase, and are only marginally resolved for the ULIRG densities.    Despite this, we show in  Appendix \ref{appendix:A} that the global properties of the disc structure and outflows are converged with respect to numerical resolution.

To better quantify the phase structure of the disc produced by SNe,   Figure~\ref{fig:phase_distr} shows volume weighted probability distribution functions (PDFs) for the gas density and temperature in both the FX and SC simulations.    The probability functions are measured in the regions $|z| < 1.5z_{\rm eff}$ (solid lines) and $|z|<50$ pc (dashed lines), and are time averaged over $30 \hbox{ Myr} < t < 40 \hbox{ Myr}$.  
The PDFs within $|z|<50$ pc describe the phase structure of the bulk of the disc while those for $|z| < 1.5z_{\rm eff}$ (solid lines) include some of the corona. 
Figure~\ref{fig:phase_distr} shows that all of the models are dominated by cool gas ($T\sim 10^4$ K) with a density close to the mean density of the disc model.   In the density probability distribution there is a low density tail associated with SNRs that experienced cooling losses and are in the momentum conserving phase.  This tail to the PDF is also present at high temperatures, though it is somewhat less prominent.

To further quantify the phase structure, Table~\ref{tab:volfill} gives the volume and mass filling fractions for low density  ($n_{\rm H}<0.01\bar{n}_{\rm H}$) and high temperature  ($T>10^{5}$ K) gas in the region $|z|<50$ pc, i.e., the region dominated by the disc.  Table~\ref{tab:volfill} shows that the low density and/or hot gas always constitutes a negligible fraction of the total mass budget of the disc. 
However, this gas occupies a reasonable fraction of the volume (up to $\sim 10 \%$), particularly for the higher surface density FX models.   The volume filling fractions of hot gas are particularly low for the SC models because of the stronger radiative losses when SNe explode in dense gas.\footnote{The volume filling fractions of low density gas are reasonably well converged and do not change  for our higher resolution `-L9' simulations; however, the volume filling fractions of high temperature gas in Table~\ref{tab:volfill} increase by a factor of a few in the higher resolution SC simulations.} 
More generally, the volume filling fractions of low density gas are larger than those of hot gas, particularly for the SC simulations in which the SN rate is proportional to the local gas density.    This again corresponds to the fact that SNRs spend  more time  in the momentum conserving phase, when radiative losses have sapped their thermal energy. 

\begin{table*}
\centering
{\bfseries Volume and mass filling fractions for $|z| < 50$ pc} 
\makebox[\linewidth]{\scriptsize
\begin{tabular}{llcccc}
\hline

& Simulation name & $f_{\rm V} (n_{\rm H}<0.01\bar{n}_{\rm H})$ & $f_{\rm V} (T>10^{5}$ K$)$ & $f_{\rm M} (n_{\rm H}<0.01\bar{n}_{\rm H})$ & $f_{\rm M} (T>10^{5}$ K$)$ \\ 
\hline
& FX-MW-L8 & $1.8\times 10^{-2}$ & $2.4\times 10^{-3}$ & $2.2\times 10^{-5}$ & $1.3\times 10^{-6}$ \\
& FX-ULTRA-MW-L8 & $3.1\times 10^{-2}$ & $1.1\times 10^{-2}$ & $5.7\times 10^{-5}$ & $3.0\times 10^{-5}$ \\
& FX-ULIRG-L8 & $7.1\times 10^{-2}$ & $4.1\times 10^{-2}$ & $1.4\times 10^{-4}$ & $3.4\times 10^{-5}$ \\
& SC-MW-L8 & $1.9\times 10^{-2}$ & $1.1\times 10^{-3}$ & $2.3\times 10^{-5}$ & $2.3\times 10^{-5}$ \\
& SC-ULTRA-MW-L8 & $2.8\times 10^{-2}$ & $5.6\times 10^{-4}$ & $5.4\times 10^{-5}$ & $1.1 \times 10^{-5}$ \\
& SC-ULIRG-L8 & $5.8\times 10^{-3}$ & $1.6\times 10^{-5}$ & $1.6\times 10^{-5}$ & $2.0 \times 10^{-7}$ \\
\hline
\end{tabular}}
\caption{ Volume and mass filling fractions of  low density/hot material; all fractions are measured in the region $|z| < 50$ pc.
Column1: simulation label. Column 2: volume filling fraction of  low density material. Column 3: volume filling fraction of  hot material. Column 4: mass filling fraction of  low 
density material. Column 5: mass filling fraction of  hot material. }\label{tab:volfill}
\end{table*}

The volume filling fraction of low density gas in our simulations is of order that expected from analytical estimates.   For example, \citet{1977ApJ...218..148M} find that the volume filling fraction of SNRs when they reach pressure equilibrium with the external medium at the end of their evolution is
\begin{equation}
f_{\rm V}=1-\exp(-Q_{\rm SNR})
\label{eq:fV}
\end{equation}
where \begin{equation}
Q_{\rm SNR} \simeq 5 \left(\frac{\dot{n}_{\rm SN}}{10^3 \hbox{ kpc}^{-3}\hbox{ Myr}^{-1}}\right)
{n}_{\rm H}^{-0.14}\tilde{P}_{04}^{-1.30}
\label{eq:QSNR}
\end{equation}
is the porosity of the ISM, $\bar{n}_{\rm H}$ is the mean 
hydrogen density, and $\tilde{P}_{04} = P/(10^4 \, {\rm k_B} \, {\rm cm^{-3} \, K})$ is a rescaled ISM pressure.
For example, for our ULTRA-MW run, $\bar n_H \sim 10 \, {\rm cm^{-3}}$, $\tilde{P}_{04} \sim 10$, and so equations \ref{eq:fV} \& \ref{eq:QSNR} imply $f_{\rm V} \sim 0.2$.  This is of order, though a bit larger than, what we find in the simulation for the volume filling fraction of low density gas.  

\subsection{The Global Dynamics of the Corona}

\begin{figure*}
    \includegraphics[width=0.95\textwidth]{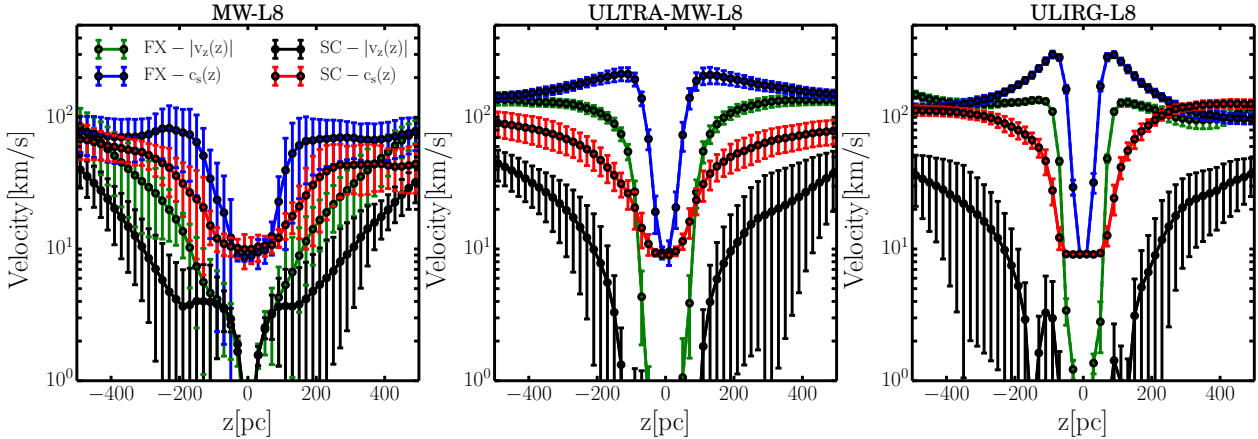}
\caption{ Profiles of the $z$-component of the velocity (green) and sound speed (blue) for simulations with spatially random SN rates in a fixed volume near the disc (FX simulations) and the $z$-component of velocity (black) and sound speed (red) for simulations with SN rates $\propto \rho$ (SC simulations). The profiles are time averaged over the interval $30 \hbox{ Myr} <t< 40 \hbox{ Myr}$ and the error bars represent the 1-$\sigma$ variations during the considered time interval.   The outflow is subsonic throughout the domain, in contrast to the standard expectation for SNe-driven galactic winds (see \S \ref{sec:bc_dep} for details).}\label{fig:prof_fx}
\end{figure*}

Figure~\ref{fig:prof_fx}  compares vertical profiles of the volume averaged vertical ($z$) component of the velocity to the sound speed for the FX (dashed lines) and SC (solid lines) simulations.  The error bars represent 1-$\sigma$ temporal variations.   These profiles have been averaged over the interval $30 \hbox{ Myr}<t<40 \hbox{ Myr}$, when ULTRA-MW and ULIRG develop a quasi-steady wind.
Figure~\ref{fig:prof_fx} shows clearly the hot corona in all of the simulations: the mean temperature at high $|z|$ is a factor of $\sim 100$ larger than that in the midplane.    In addition, all of our simulations show acceleration as the gas moves from the mid-plane towards the box boundary.  However, all of the gas motions remain subsonic on the computational domain, with outflow velocities at most $\sim 100 \, {\rm km \, s^{-1}}$.   In detail, the velocity structure  is sensitive to the SN seeding scheme, with the SC runs typically having slower winds. {This is consistent with the results of the previous subsections, which showed that SNe in the SC model are more susceptive to radiative losses. The subsonic outflows seen here are, however, contrary to expectations for SNe-driven galactic winds (e.g., \citealt{CC85}). 
As we will emphasize in \S \ref{sec:bc_dep}, this result is a generic limitation of local Cartesian simulations.}

\section{Energetics \& Outflow Properties}
\label{sec:energetics}

In this section, we analyse the state of the stratified simulations in terms of the balance between the injection of energy by SNe and  losses via cooling and  outflows. In steady state, the equation for energy balance is
\begin{equation}
\dot{E}_{\rm SN}=\dot{E}_{\rm cool}+\dot{E}_{\rm out},
\label{eq:energy}
\end{equation}
where $\dot{E}_{\rm SN}\geq 0$ is the energy injection rate due to SNe,  $\dot{E}_{\rm cool}\geq0$ is the cooling rate, and $\dot{E}_{\rm out}\geq 0$ is the rate at which the wind carries away energy.   In addition to these sources and sinks, we also quantify $\dot{E}_{\rm turb}\geq 0$, the rate at which turbulent energy cascades to small scales.   This quantifies how efficient SNe are at driving turbulence, but does not show up as a separate term in equation \ref{eq:energy} because the energy in turbulent motions ultimately goes into heat and thus radiation.   

In many ways, it is more useful to analyze the properties of turbulence driven by stellar feedback in terms of the momentum supplied by SNe and other sources (\S \ref{sec:turb}; see, e.g., \citealt{2005ApJ...630..167T, 2011ApJ...731...41O}).   However, for the purposes of understanding SNe-driven galactic winds, it is particularly instructive to quantify the fraction of SNe energy that goes into the wind.  This and the wind mass loading are in fact the two key parameters of SNe-driven galactic wind theory (e.g., \citealt{CC85}).   For this reason we focus in this section on analyzing the energetics and outflow mass loading in our local simulations.  

\subsection{Definitions}\label{sec:energy_defs}

To quantify the cooling rate, we directly adopt the algorithm used in {\sc ramses} to compute the cooling rate $\dot{E}_{\rm cool}$ given the density and temperature distribution at a given time. We measure the cooling rate in the region $|z|<1.5z_{\rm eff}$ which is larger than the steady state disc scale height and where most of the cooling losses are achieved. 

We define two energy injection rates associated with SNe.    The total SN energy injection rate in the case of infinite resolution is $\dot{E}_{\rm SN,tot}(t)=\dot{n}_{\rm SN}(t)\times 10^{51} \hbox{erg}$.   In practice, however, our sub-grid model (\S \ref{sec:SN_seeding}) takes into account that radiative losses on unresolved scales sap the SNR of energy.   We thus separately define the energy injection rate using the sub-grid model as $\dot{E}_{\rm SN}(t)$. 
The ratio $\dot{E}_{\rm SN}/\dot{E}_{\rm SN,tot}$  quantifies how well our simulations resolve the energy-conserving Sedov-Taylor phase of SNRs: a ratio near unity means that the SNe typically explode in regions where the density is low enough to allow the simulation to resolve the full SNR evolution.  When the ratio is low, by contrast, the SNe seeding   injects primarily the residual energy/momentum after cooling losses have already modified the evolution of the SNR. The MW simulations naturally have low density (large cooling radii) and $\dot{E}_{\rm SN}/\dot{E}_{\rm SN,tot} \approx 1$, consistent with fully resolved SNRs, however the ratio is $<1$ for ULTRA-MW and ULIRG where the typical densities are higher and the sub-grid approach is required.  

To quantify the mass outflow rate $\dot{M}_{\rm out}$ from the domain, we compare it to the rate at which stars would be formed in the model disc. We do so by converting the 
SN rate $\dot{n}_{\rm SN}$ into a star formation rate ${\rm SFR} = \int dz  L^2 m_*\dot{n}_{\rm SN}$ with $m_* = 100$ M$_\odot$ and   defining the mass loading factor of the wind as
$\eta=\dot{M}_{\rm out}/{\rm SFR}$.   To quantify the outflow of energy from a given region, consider a slab with the mid-plane at its centre whose upper and lower boundaries are defined by $|z|=z_{\rm out}$. We compute the outflow rate as the following sum:
\begin{equation}\label{eq:outflow}
\dot{E}_{\rm out}=\sum_{\rm i, |z_{\rm i}|=z_{\rm out}} \rho_{\rm i} v_{\rm z,i} \Delta x^2 \left(v_{\rm i}^2+\frac{\gamma}{\gamma-1}\frac{P_{\rm i}}{\rho_{\rm i}}\right),
\end{equation}
where $\rho_{\rm i}$ is the cell density, $v_{\rm z,i}$ is the z-component of the  velocity, $v_{\rm i}$ is the magnitude of the  velocity  and $P_{\rm i}$ is the  pressure.  Only cells with outflowing z-velocity are included in the sum.   We return later to the fact that most of the outflowing material at a given height falls back eventually and only a small amount leaves the box. 
Equation \ref{eq:outflow} corresponds to the Bernoulli parameter of the outflow times the mass flux, except that we do not include a term with the gravitational potential energy.   The reason is that the local boxes we consider  do not have a well defined escape velocity. In fact, $v_{\rm esc}\propto z$  for large $z$.    

Finally, the rate at which SNe energy goes into turbulent motions can be estimated if some of the properties of turbulence are known.  For the turbulence analysis we decompose the velocity field into a mean and fluctuating component
\begin{equation}
 {\bf v}(x,y,z)={\bf v}_{\rm bg}(z)+ {\bf \delta v}(x,y,z)
\end{equation}
where ${\bf v}_{\rm bg}(z)$ is the background value at height $z$ and ${\bf \delta v}$ is the velocity fluctuation defined such that $\langle {\bf \delta v} \rangle_z = 0$ where $\langle \rangle_z$ denotes an average over $x, y$ at a given height z.   Given the velocity fluctuations ${\bf \delta v}$, we calculate the 1D mass (volume) weighted 
velocity dispersion $\sigma_{\rm v,M}=\left\langle \delta v^2/3\right\rangle_{\rm M}^{1/2}$ ($\sigma_{\rm v,V}=\left\langle \delta v^2/3\right\rangle_{\rm V}^{1/2}$), and the turbulent kinetic energy density $\epsilon_{\rm kin}=\left\langle \rho \delta v^2\right\rangle/2$ (where $\langle \rangle$ now denotes a full volume average). 
We also measure $E(k)$, the 1D power spectrum of the turbulent kinetic energy density: we first take the power spectra of the $x,y,z$-components of the kinetic energy density separately and then sum the contributions. 
We can then define an effective driving scale $L_{\rm drive}$ for the turbulent motions:
\begin{equation}
 L_{\rm drive}=\frac{\int_{k_{\rm min}}^{k_{\rm max}} \frac{2\pi}{k}E(k)dk}{\int_{k_{\rm min}}^{k_{\rm max}} E(k)dk}.
 \label{eq:Ldrive}
\end{equation}
The minimum wave number $k_{\rm min}$ is determined by the maximum wavelength that can be sampled 
$\lambda_{\rm max} = L/2$, i.e. $k_{\rm min}=2\pi/\lambda_{\rm max}$. The maximum wave number depends on the spatial resolution, $k_{\rm max}=2\pi /\Delta x$.    Note that because the turbulent kinetic energy is highly concentrated in the disc, the exact value of $L_{\rm drive}$ is insensitive to the details of the coronal dynamics, but is instead determined by the bulk of the mass, which resides in the disc.

In absence of continuous driving by SNe, turbulence is expected to decay on a time scale $t_{\rm decay}$ that can be approximated by (e.g., \citealt{Stone1998})
\begin{equation}
 t_{\rm decay}\approx\frac{L_{\rm drive}}{\sigma_{v,M}}.
\end{equation}
{In our simulations, we measure $\sigma_{v,M}$ by averaging over the region $|z|<1.5z_{\rm eff}$, which should include the contribution to the velocity dispersion from turbulent eddies of all scales relevant to disc turbulence since $3z_{\rm eff} \gtrsim 2L_{\rm drive}$ for all our simulations (Table~\ref{tab:turb}).} 
We can thus estimate the rate of turbulence {driving} by the rate which the turbulent energy would decay absent driving:
\begin{equation}
 \dot{E}_{\rm turb}\approx\frac{\epsilon_{\rm kin}L^3}{t_{\rm decay}}=\left\langle \frac{1}{2}\rho \delta v^2\right\rangle \frac{\sigma_{v,M}}{L_{\rm drive}}L^3
 \label{eq:edotturb}
\end{equation}
Note that because of approximations inherent in deriving equation \ref{eq:edotturb}, $\dot E_{\rm turb}$ is only  accurate to multiplicative factors of order unity (in contrast to $\dot E_{\rm cool}$,  $\dot E_{\rm SN}$, and $\dot E_{\rm out}$). 

\subsection{Results of the Energetics Analysis}\label{sec:energetic}

\begin{figure*}
    \includegraphics[width=0.99\textwidth]{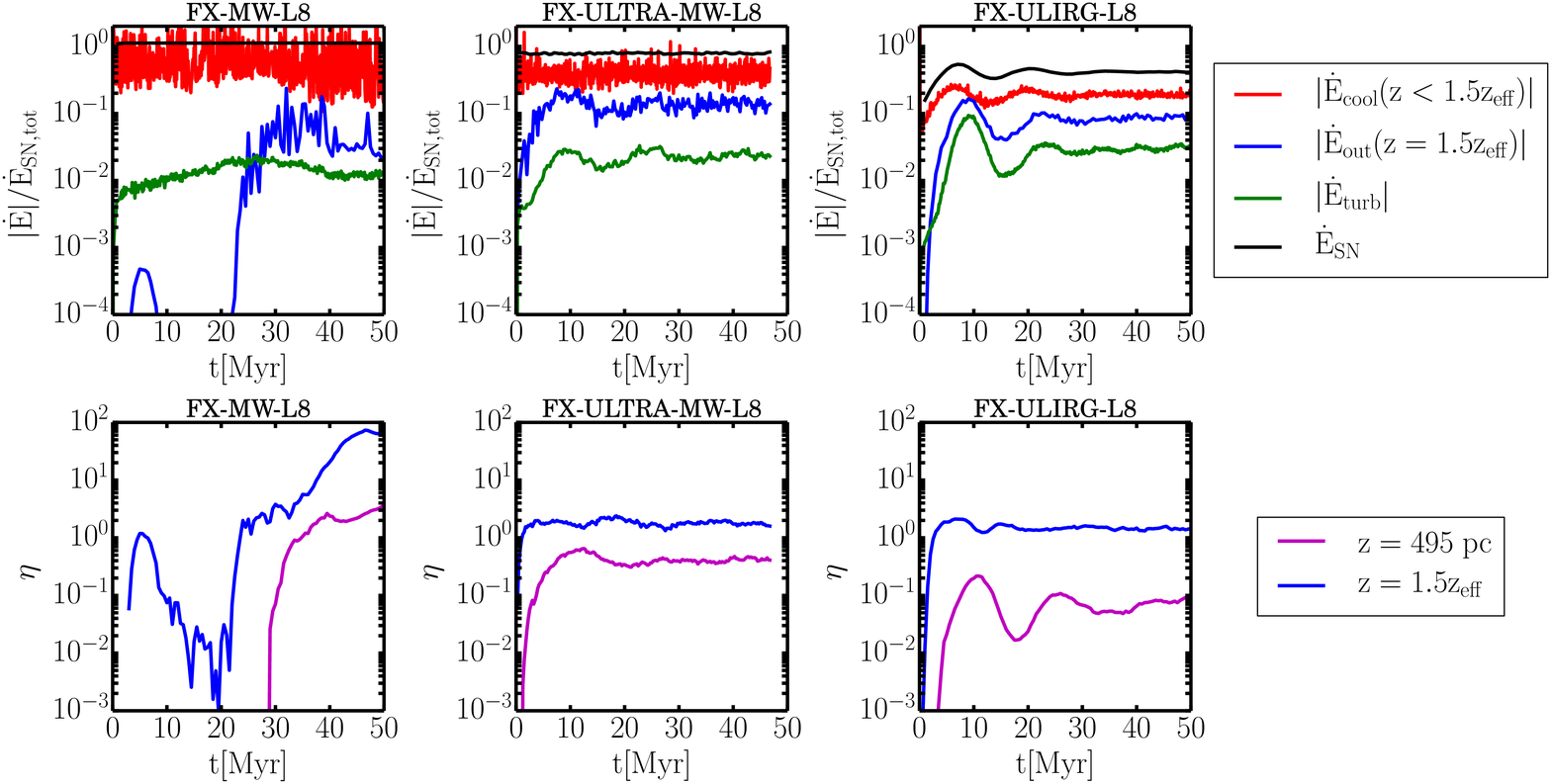}
\caption{Energetics and outflow mass loading for the simulations with SNe random in space (FX simulations).  First row: SN energy injection rate given our subgrid SN model (black), cooling rate 
(red), outflow rate from the region near the disc, $|z|<1.5 z_{\rm eff}$(blue), and turbulence driving rate in the disc (green).  All $\dot E$ are normalized by the total energy injection rate by SNe ($10^{51}$ erg per SN).   Second row: wind mass loading factor  $\eta$ (ratio of outflow rate to star formation rate) at $z=1.5 z_{\rm eff}$ (blue) and 
at $z=495$ pc (magenta).    The outflow carries away at most $\sim 10 \%$ of the SN energy while the mass loading factor decreases with increasing height $z$. 
Figure \ref{fig:energetics_sc} shows the same results for the SC runs in which the SNe rate $\propto \rho$.   The FX models shown here drive more powerful winds with more energy and larger 
mass loading.}\label{fig:energetics_fx}
\end{figure*}

\begin{figure*}
    \includegraphics[width=0.99\textwidth]{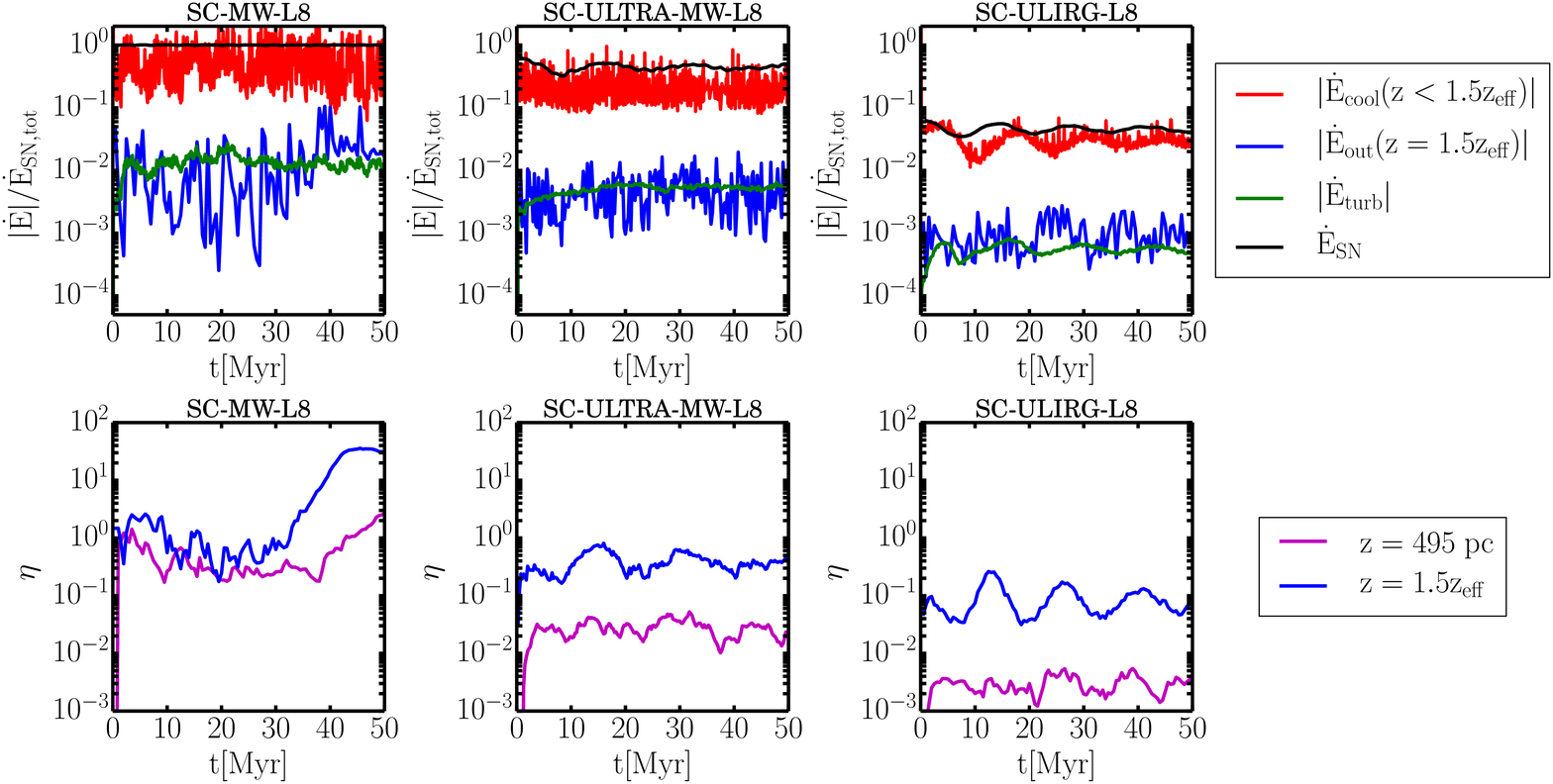}
\caption{ Energetics and outflow mass loading for the simulations with SNe rates $\propto \rho$ (SC simulations).  First row: SN energy injection rate given our subgrid SN model (black), cooling rate 
(red), outflow rate from the region near the disc, $|z|<1.5 z_{\rm eff}$(blue), and turbulence driving rate in the disc (green).  All $\dot E$ are normalized by the total energy injection rate by SNe ($10^{51}$ erg per SN).   Second row: wind mass loading factor $\eta$ (ratio of outflow rate to star formation rate) at $z=1.5 z_{\rm eff}$ (blue) and 
at $z=495$ pc (magenta).    Figure~\ref{fig:energetics_fx}  shows the same results for the FX runs in which the SNe are random in space.   The SC model is significantly less efficient than the FX model at driving winds with $\dot E$ and mass loading factors $\eta$ smaller by a factor of $\sim 10$.  }\label{fig:energetics_sc}
\end{figure*}

\begin{figure*}
    \includegraphics[width=0.99\textwidth]{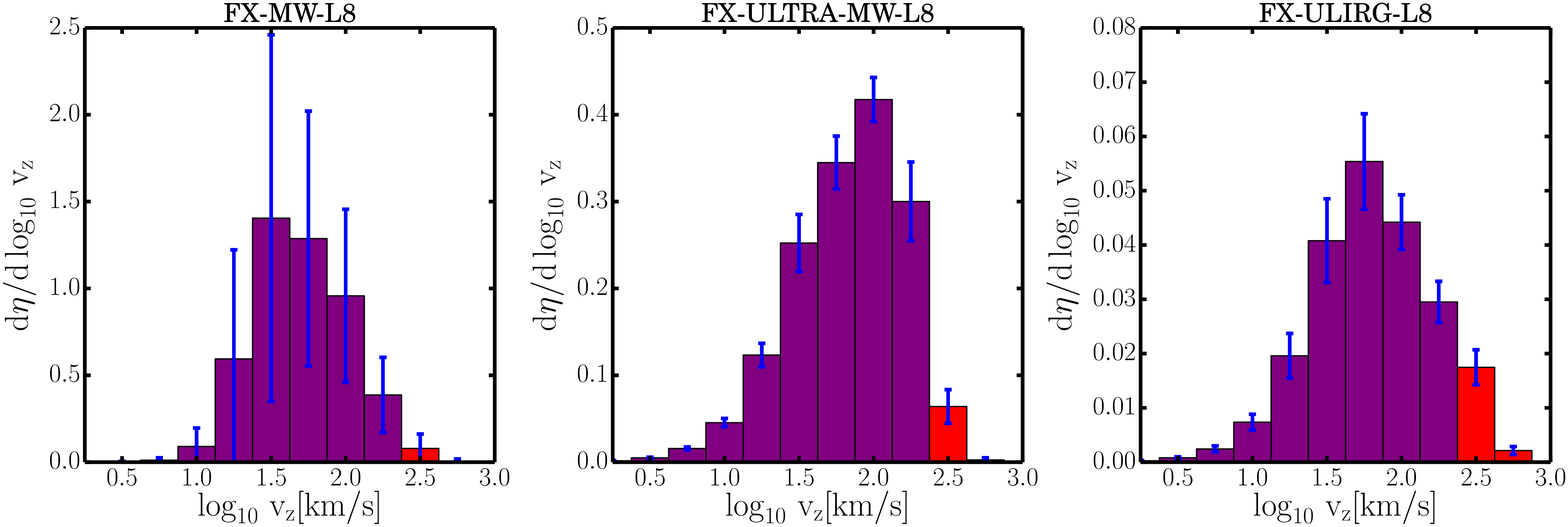}
\caption{ Wind mass loading factor $\eta$ (ratio of outflow rate to star formation rate) at the edge of the computational domain ($|z|=495$ pc) as a function of  $\log_{\rm 10} v_{\rm z}$ for the simulations with SNe random in space (FX simulations).  These results are time averaged over $30 \hbox{ Myr} <t< 40 \hbox{ Myr}$. The error bars represent the 1-$\sigma$ 
time variation in the considered time interval.  Outflow with $|v_{\rm z}| \geq 300$ km/s is represented by the red vertical bars.  Very little of the outflow is moving sufficiently fast to escape a galaxy with an escape velocity typical for galaxies with the star formation rate surface densities modeled here.}\label{fig:mass_loading_pdf}
\end{figure*}

Figures~\ref{fig:energetics_fx} (FX-runs) and \ref{fig:energetics_sc} (SC-runs) summarise the energetics and wind mass loading factors 
for our  three different galaxy simulations.   All of the $\dot E$ values are normalized to the total SNe energy injection rate $\dot E_{\rm SN,tot}$ for the given simulation.    Appendix \ref{appendix:A} shows that the global quantities  in Figures~\ref{fig:energetics_fx} (FX-runs) and \ref{fig:energetics_sc} (SC-runs) are well converged with respect to spatial resolution.    The key results of Figures~\ref{fig:energetics_fx} (FX-runs) and \ref{fig:energetics_sc} are as follows:
\begin{itemize}
\item  $\dot{E}_{\rm SN}/\dot{E}_{\rm SN,tot}$ is  highest in the MW case where the SNR evolution is easier to resolve and  lowest in the ULIRG case.   $\dot{E}_{\rm SN}/\dot{E}_{\rm SN,tot}$ is also systematically higher in the FX simulations relative to the SC simulations.  This is because in the FX simulations  the SN rate per unit volume is independent of height for $|z| < z_{\rm eff}$; more SNe thus explode in low density material where SNRs easier to resolve.

\item   In all cases the vast majority of the energy supplied by SNe is radiated away.  The turbulence driving rate is typically $\dot{E}_{\rm turb}\lesssim 10^{-2}\dot{E}_{\rm SN,tot}$; we derive this analytically in the next section.   
Outflows carry away at most $\sim 10 \%$ of the total energy supplied by SNe.   This high outflow rate of energy is achieved only in the FX simulations, because in that case SNe couple better to the low density material.   By contrast, in the SC simulations the outflows carry away at least a factor of $\sim 10$ less energy, $\lesssim 1 \%$ of the total energy supplied by SNe.   This is consistent with the much cooler atmosphere in Figure \ref{fig:maps_sc} relative to Figure \ref{fig:maps_fx}.

\item In the FX simulations, the relative power in the outflows $\dot{E}_{\rm out}/\dot{E}_{\rm SN,tot}$ is nearly the same in ULTRA-MW and ULIRG (which have more clearly established steady outflows than the MW simulation).   By contrast, in the SC simulations, the relative power in the outflows $\dot{E}_{\rm out}/\dot{E}_{\rm SN,tot}$ declines significantly with increasing surface density of the disc.

\item The mass loading factor of the wind $\eta$ decreases with increasing gas surface density and with increasing height $z$.  The dependence on height $z$ can be understood using Figure~\ref{fig:mass_loading_pdf}, which shows a histogram of $d\eta / d\log_{\rm 10} v_{\rm z}$, the contribution to the mass loading  at $|z| = 495$ pc (the edge of the domain) from different logarithmic bins in $\log_{\rm 10} v_{\rm z}$.   Figure~\ref{fig:mass_loading_pdf} shows that the vast majority of the outflow is at a relatively low velocity that {would not escape a realistic galactic potential} appropriate to these high star formation rate surface densities.   For example,  $\eta (v_{\rm z}>300 \hbox{ km/s})=$ 0.02, 0.016 and 0.005 for FX-MW-L8, FX-ULTRA-MW-L8 and FX-ULIRG-L8, respectively.    The low mass loading of high velocity gas explains why $\eta$ declines significantly with increasing height, since much of the material does not have sufficient energy to reach large $z$.   This property of the local simulations precludes us from robustly predicting the outflow rates associated with a given set of galaxy properties (e.g., surface density and SN rate).   We return to this  in \S \ref{sec:bc_dep} (see also Appendices \ref{appendix:B} \& \ref{appendix:C}).

\end{itemize}

\section{Properties of Supernova-driven Disc Turbulence}\label{sec:turb}
\subsection{Disc Turbulence}
Table~\ref{tab:turb} summarises the statistical properties of the turbulence in steady state for our various simulations, including the mass and volume weighted turbulent velocity dispersions and the characteristic driving scale of the turbulence $L_{\rm drive}$ (eq. \ref{eq:Ldrive}).  These quantities are all measured by averaging over the region $|z|<1.5z_{\rm eff}$ and thus largely correspond to the properties of the turbulence in the disc, not the outflow (this is particularly true for the mass-weighted quantities; volume-weighted turbulent velocity dispersions depend somewhat more on the size of the region over which the average occurs).  One of the key results in Table ~\ref{tab:turb} is that the mass weighted velocity dispersion is $\approx 3-7 \, {\rm km \, s^{-1}}$, with little dependence on gas surface density or the SN driving model (analogous results have been reported in the numerical work of \citealt{2006ApJ...638..797D}, \citealt{2009ApJ...704..137J}, \citealt{2012ApJ...754....2S}, \citealt{2013ApJ...776....1K}).  The volume weighted velocity dispersion is, not surprisingly, significantly larger than the mass-weighted dispersion, but also depends more sensitively on the spatial distribution of the SNe (FX vs. SC).   Finally, the effective driving scale of the turbulence $L_{\rm drive}$ is of order a few scale-heights in all cases. 
{In the Milky Way, the driving scale can be probed by measuring the correlation scale of molecular clouds with similar inclinations with respect to the Galactic disc. Using this method, \cite{2009MNRAS.398.1201D} estimated a driving scale $\gtrsim 100$ pc comparable to that found in our simulations. In external galaxies, the driving scale can be estimated by measuring the ISM velocity power spectrum, for example using HI gas observations \cite[e.g.][]{2015ApJ...810...33C}.}

\begin{table*}
\centering
{\bfseries Turbulence properties}
\makebox[\linewidth]{\scriptsize
\begin{tabular}{lccccccc}
\hline
 Label & Type & $\bar{n}_{\rm H}$ [cm$^{-3}$] & $\sigma_{\rm v,M}$ [km/s] & $\sigma_{\rm v,V}$ [km/s] & $L_{\rm drive}$ [pc] &  $\sigma_{\rm mod}$ [km/s] & $\chi$ [km/s] \\ 
\hline
FX-MW-L8 & MW & 0.7 & 5.6 & 11 & 119 & 5.4 & 160 \\
FX-ULTRA-MW-L8 & ULTRA-MW & 5.3 & 7.4 & 140 & 126 & 7.5 & 50 \\
FX-ULIRG-L8 & ULIRG & 61.8 & 7.1 & 180 & 93 & 6.2 & 94\\
SC-MW-L8 & MW & 0.7 & 6.4 & 14 & 138 & 6.8 & 93 \\
SC-ULTRA-MW-L8 & ULTRA-MW & 5.4 & 6.4 & 33 & 62 & 6.9 & 34 \\
SC-ULIRG-L8 & ULIRG & 63.1 & 3.0 & 30 & 54 & 4.2 & 30 \\
\hline
\end{tabular}}
\caption{Properties of the turbulence in the statistical steady state in the region near the disc, $|z|<1.5z_{\rm eff}$. Columns are:  simulation label, type of disc initial conditions, mean gas density, mass weighted 1D velocity dispersion, 
volume weighted 1D velocity dispersion, effective driving scale (eq. \ref{eq:Ldrive}), velocity dispersion predicted by eq. \ref{eq:sigma_martizzi}, $\chi$ parameter defined in eq.~\ref{eq:chi} and used in \S \ref{sec:SF}.}\label{tab:turb}
\end{table*}

\cite{2015MNRAS.450..504M} derived  a simple estimate of the turbulent velocity dispersion produced by SNe feedback as a function of mean ambient density and volumetric SN rate, and showed that it agreed with simulations of multiple SNe in a periodic box.  Their result was
\begin{equation}
\sigma_{\rm mod}\approx \frac{3}{4\pi} \left( \frac{32 \pi^2}{9} \right)^{3/7}\left[ \frac{P_{\rm fin}(\bar{n}_{\rm H})}{\bar{\rho}}\right]^{4/7} \left( f \, \dot{n}_{\rm SN} \right)^{3/7},
\label{eq:sigma_martizzi}
\end{equation}
where $P_{\rm fin}(\bar{n}_{\rm H})$ is the average momentum injected per SN in the momentum-driven snow-plough phase in a medium of mean density $\bar{n}_{\rm H}$; 
$\bar{\rho}$ is the corresponding mass density; and $f$ is a factor that takes into account momentum cancellation when multiple blast-waves interact and order unity effects related to gradients in the density and pressure of the medium.   Table~\ref{tab:turb} shows that equation \ref{eq:sigma_martizzi} does a good job of reproducing the simulated values of $\sigma_{\rm v,M}$ in our stratified media simulations, over a wide range of surface densities.    For this comparison, we used \begin{equation}\label{eq:pfin_martizzi}
 \frac{P_{\rm fin}}{m_*}=1,420~\hbox{\rm km s$^{-1}$} \left(\frac{n_{\rm H}}{100~{\rm cm}^{-3}}\right)^{-0.160}
\end{equation}
from \cite{2015MNRAS.450..504M} and assumed $f = 0.4$ and  $f=0.3$ for FX and SC, respectively. 

One of the results of the  simulations (both ours and previous work) is that the mass-weighted turbulent velocity dispersion driven by SNe does not depend strongly on the surface density of the disc.   To understand better why this is the case, and to see why this result is likely robustly applicable to a range of galaxy environments, it is helpful to recall that in feedback models,  the vertical weight of the galactic disc is balanced by the turbulent pressure generated by stellar feedback \citep[see][]{2005ApJ...630..167T, 2011ApJ...731...41O, 2013MNRAS.433.1970F}.  The turbulent pressure provided by SNe can be written as
\begin{equation}\label{eq:pturb_star}
 p_{\rm T}
 \approx F \dot{\Sigma}_*\left(\frac{P_{\rm fin}}{m_{*}}\right)
\end{equation}
where $F$ is another factor of order unity that takes into account cancellation of momentum provided by SNe in the disc, the fraction of momentum lost in driving galactic winds and the ratio of the turbulence driving scale and the disc scale height.   A second equation for $p_{\rm T}$ can be derived from vertical hydrostatic equilibrium.  For the simulations in this paper that lack self-gravity, this takes the form
\begin{equation}\label{eq:pturb}
 p_{\rm T}\approx \pi G \Sigma_{\rm gas} \Sigma_{\rm *}
\end{equation}
where $\Sigma_{\rm gas}$ is the gas surface density, $\Sigma_{\rm *} = \Sigma_{\rm gas}/f_{\rm gas}$ is the stellar surface density in the disc, and $f_{\rm gas}$ is the gas fraction. 
Equations~\ref{eq:pturb_star} and \ref{eq:pturb} yield the star formation law for discs self-regulated by SN-driven turbulence:
\begin{equation}\label{eq:KS}
\dot{\Sigma}_*\approx\frac{\pi G}{F(\frac{P_{\rm fin}}{m_*}) f_{\rm gas}}\Sigma_{\rm gas}^2.
\end{equation}
Equation~\ref{eq:KS} reduces to a Kennicutt-Schmidt law $\dot{\Sigma}_{*}\propto \Sigma_{\rm gas}^2$ if all the 
pre-factors do not depend on  surface density.  

Substituting equation~\ref{eq:KS} into equation~\ref{eq:sigma_martizzi}, using the fact that (i) the SN rate is $\dot{n}_{\rm SN}\approx\dot{\Sigma}_*/2 h m_*$, where $h=\sigma_{\rm mod}^2/2 \pi G \Sigma_*$ is the scale height of the disc, and (ii) the average density in the disc can be approximated by $\bar{\rho}\approx \Sigma_{\rm gas}/2 h$, we find
\begin{align}\label{eq:sigma_weak_function}
\sigma_{\rm mod}\approx 1.8 \left(\frac{f}{F}\right)^{3/5} G^{2/5} P_{\rm fin}^{1/5}f_{\rm gas}^{-2/5}\Sigma_{\rm gas}^{1/5}.
\end{align}
Equation \ref{eq:sigma_weak_function} shows explicitly that the turbulent velocity is predicted to be a very weak function of disc properties, thus explaining the weak dependence of velocity dispersion on gas surface density found in our (and previous) numerical simulations (see Table \ref{tab:turb}). 

The simple estimates in this section can also explain why the energy input into turbulence $\dot{E}_{\rm turb} 
\sim 10^{-2}\dot{E}_{\rm SN,tot}$ (Figs \ref{fig:energetics_fx} \& \ref{fig:energetics_sc}).   This follows from  $\dot{E}_{\rm turb}/\dot{E}_{\rm SN,tot}  \sim \dot{P}_{\rm in}\sigma_{\rm v,M}/\dot{E}_{\rm SN,tot}\approx P_{\rm fin}\sigma_{\rm v,M}/10^{51}$ erg, where $\dot{P}_{\rm in}$ is the momentum injection rate. Since $P_{\rm fin} \sim100 \hbox{ M}_{\odot} \times 1,000 \hbox{ km/s}$ and $\sigma_{\rm v,M}\sim10$ km/s,  $\dot{E}_{\rm turb}/\dot{E}_{\rm SN,tot} \sim 10^{-2}$.   In addition, the weak dependence of $P_{\rm fin}$ and $\sigma_{\rm v,M}$ on gas surface density leads to the weak dependence of $\dot{E}_{\rm turb}/\dot{E}_{\rm SN,tot}$ on galaxy model seen in Figures \ref{fig:energetics_fx} \& \ref{fig:energetics_sc}.

\subsection{Implications for the Star Formation Law}
\label{sec:SF}
Equation~\ref{eq:KS} provides a simple analytical estimate of the star formation law in gaseous discs. To test the assumptions of this model using our simulations, we use the parameter
\begin{equation}
 \chi=\frac{f_{\rm gas}\bar{\rho}\sigma_{\rm v,M}^2}{\dot{\Sigma}_*} = F \left(\frac{P_{\rm fin}}{m_{*}}\right)f_{\rm gas}
 \label{eq:chi}
\end{equation}
where the latter equality follows from equation~\ref{eq:KS}.    Alternatively, equation~\ref{eq:KS} can be rewritten as $ \dot{\Sigma}_*\approx \pi G \chi^{-1} \Sigma_{\rm gas}^2$.
The dependence of $\chi$ on $\Sigma_{\rm gas}$ thus quantifies the tilt in the Kennicutt-Schmidt law with respect to $\dot{\Sigma}_{*}\propto \Sigma_{\rm gas}^2$.   Table~\ref{tab:turb} summarises the values of $\chi$ in our simulations (we calculate $\chi$ directly using the first definition in equation \ref{eq:chi}).   Table~\ref{tab:turb} shows that in our calculations, $\chi$ does not strongly depend on the gas surface density.  Our results are thus consistent with a Kennicutt-Schmidt law close to $\dot{\Sigma}_* \propto \Sigma_{\rm gas}^2$ \citep[as in previous numerical work in local simulations;][]{2012ApJ...754....2S, 2013ApJ...776....1K}.    We leave a detailed study of the star formation law to future work.


\section{Wind Dynamics:  Effects of Box Size and Geometry}\label{sec:bc_dep}

Appendix~\ref{appendix:A} shows that the main properties of the SN-driven turbulence and winds converge well with increasing resolution.    Nonetheless, we believe that many aspects of realistic galactic wind dynamics are {\em not} well described by  local simulations (ours and previous work) because of limitations of the Cartesian geometry and the lack of a well-defined escape velocity.   We briefly enumerate these concerns here.

\begin{figure}
    \includegraphics[width=0.49\textwidth]{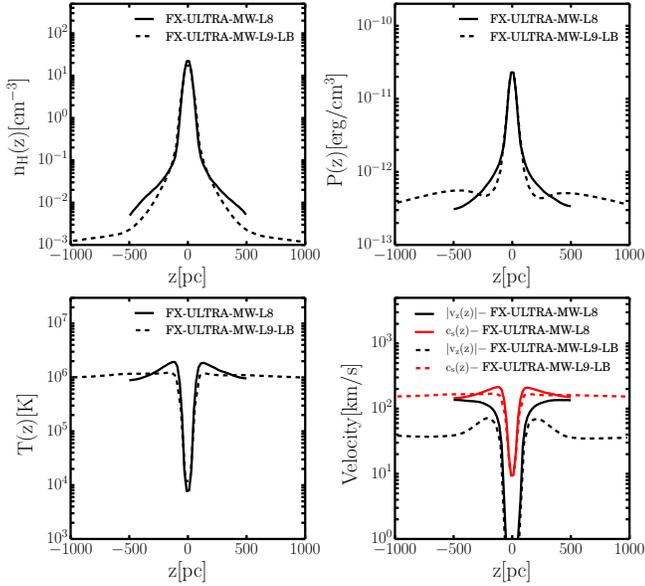}
\caption{Effect of box size on the time averaged vertical profiles (over $30 \hbox{Myr}<t<40 \hbox{Myr}$) of various quantities, for the ULTRA-MW model.  The solid line is the fiducial simulation with a 1 kpc$^3$ box (FX-ULTRA-MW-L8) while the dashed line is for a 8 kpc$^{3}$ box, i.e., each side is 2 kpc (FX-ULTRA-MW-L9-LB). Top left: density. Top-right: pressure. Bottom left: temperature. Bottom right: sound speed (red) and $z$-component of the velocity (black).  The two simulations agree near the disc midplane but the profiles differ above $|z| \sim 200$ pc.  Note, in particular, that the outflow is significantly more subsonic in the larger box.   }\label{fig:prof_lb}
\end{figure}

We compare our fiducial run FX-ULTRA-MW-L8 to FX-ULTRA-MW-L9-LB, which is the same simulation performed in a cubic box whose sides are larger by a factor 2.   Figure~\ref{fig:prof_lb} compares the density, pressure, temperature and velocity profiles time averaged within the time interval $30~\hbox{Myr}<t<40~\hbox{Myr}$.   Figure~\ref{fig:prof_lb} shows that the properties of the disc within $-200~\hbox{pc}<z<+200~\hbox{pc}$ are not influenced by the size of the domain. On the other hand, the structure of the wind/atmosphere is affected. In the large box the density profile declines to lower values and the pressure profile is much flatter; the temperature of the atmosphere is very similar, though the precise profiles differ. In both cases the velocity of the wind is subsonic throughout the  box. However in the large box the velocity is significantly more subsonic such that wind velocity close to the box boundary is a factor $\sim 2$ smaller in the large box. These differences indicate that {the detailed structure of the outflows in local Cartesian simulations of stratified discs depends on the size of the computational domain.} This result is somewhat different from \cite{2013MNRAS.429.1922C}, who found a weaker dependence on the boundary conditions. However their tests only focus on the first 10 Myr of evolution, which is less than one dynamical time of the disc.   We find that significant differences in the profiles are only seen for $t>20$ Myr, when the flow reaches its statistically steady state structure.   

\begin{figure}
    \includegraphics[width=0.49\textwidth]{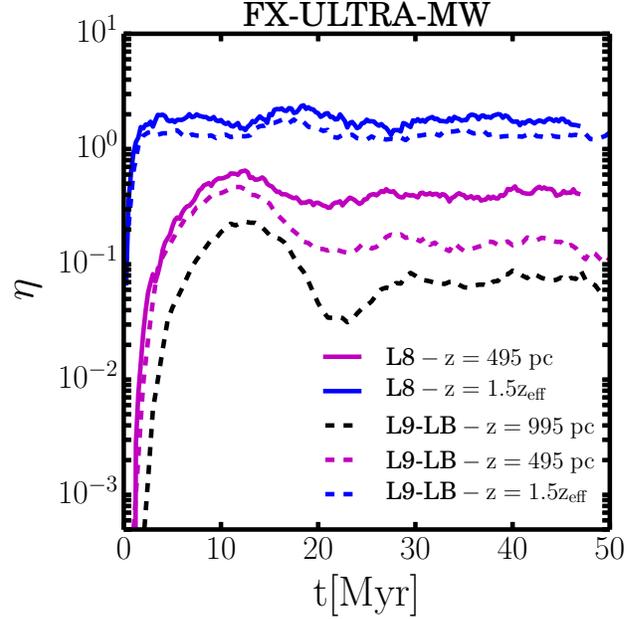}
\caption{Effect of box size on the wind mass loading factor $\eta$ (the ratio of outflow rate to star formation rate) at different heights, for the ULTRA-MW model.  We compare the fiducial simulation with a 1 kpc$^3$ box (FX-ULTRA-MW-L8) to one with an 8 kpc$^{3}$ box, i.e., in which each side is 2 kpc (FX-ULTRA-MW-L9-LB).   The two simulations agree reasonably well at $z = 1.5 z_{\rm eff} \simeq 150$ pc.  However, the mass outflow rate continues to decrease with increasing height $z$ in the larger box, indicating that there is no well-defined (i.e., independent of box properties) outflow rate in these local Cartesian simulations.   We confirm this with even taller boxes in Appendix \ref{appendix:B} and discuss the reasons for this in \S \ref{sec:bc_dep}. }\label{fig:energetics_lb}
\end{figure}

Figure~\ref{fig:energetics_lb}  compares the wind mass loading factors at several heights in
FX-ULTRA-MW-L8 and FX-ULTRA-MW-L9-LB.    The mass loading factor is very similar in the 2 runs at $z=1.5z_{\rm eff}$, but differs by a factor 2 at $z=495$ pc.  In addition, the simulation with the large box shows that the mass loading factor continues to decrease with increasing distance from the mid-plane (see the black dashed line; $\eta(z=995 \hbox{pc})$).  
In Appendix \ref{appendix:B}, we show simulations with `tall boxes' having a vertical extent 5 times larger than their horizontal extent.   The mass loading factor continues to decrease with increasing height in the taller boxes (Fig. \ref{fig:energetics_athena}).   As a result, {there is no well-defined mass outflow rate (or energy loss rate) in these local Cartesian simulations of SNe-driven galactic winds.}   Analogous concerns have been raised in the context of shearing box simulations of accretion discs, albeit with quite different physics (e.g., \citealt{Fromang2013}).   

There are likely two reasons for the lack of a well defined outflow rate in the local simulations.  The first is that the local simulations do not have a well defined escape velocity, since $v_{\rm esc} \propto z$ at large heights.   However, a more serious issue is that in a Cartesian geometry there are no adiabatic steady state winds that undergo a subsonic to supersonic transition.  We discuss this explicitly in Appendix \ref{appendix:C} and show that the standard supernovae driven winds of galactic wind theory \citep{CC85} do not exist in Cartesian geometry.   This is because of the lack of the $1/r^2$ spherical divergence that enables super-sonic winds in spherical geometry.   The absence of such wind solutions in  Cartesian geometry is presumably why the time averaged velocity profiles in all of our simulations remain mildly subsonic.    This also calls into question {\em quantitative} results on galactic winds derived from stratified Cartesian simulations.  Such quantitative analyses likely require a more realistic cylindrical or spherical geometry.\footnote{For a hypothetical sufficiently tall local Cartesian simulation, there is no outflow because the potential increases $\propto z^2$ at large $|z|$.   Such a local model would only produce a galactic fountain with no net outflow at any height $|z|$.  One could crudely estimate a global outflow rate by arbitrarily defining an escape speed $v_{\rm esc}$ motivated by a global model and calculating the mass outflow rate for $|v| > v_{\rm esc}$ at some chosen height.  In addition to depending on these arbitrary choices, the corresponding $\eta$ will be very small, as shown by Figure \ref{fig:mass_loading_pdf}.  Moreover, we believe that the resulting outflow rate would still not be accurate given the limitations of Cartesian boxes for capturing subsonic to supersonic wind transitions.}  Local Cartesian simulations are nonetheless valuable for providing insight into the conditions under which multiple supernovae generate an energetically important hot ISM, and an associated hot galactic corona.


\section{Summary and conclusions}
In this paper, we analysed a series of idealized simulations of SN feedback designed to model local Cartesian patches of vertically stratified galactic discs ($\sim 1$ kpc$^{3}$ regions).
We explored three different disc models of increasing gas surface density, from a Milky Way-like model to a ULIRG-like model. 
An intermediate model, ULTRA-MW, served as the pivot point for our analysis. 
For each disc model, we investigated two different SN seeding schemes (random in space [FX] and proportional to local gas density [SC]). 
Our primary goals were to understand how SN feedback drives turbulence in galactic discs and launches galactic winds. 
Our simulations build on our previous work \citep{2015MNRAS.450..504M} that focused on isolated SNRs and multiple SNe in periodic boxes.  In particular, we implemented the subgrid model derived in that study to capture the momentum injection and thermal energy of individual SNe when the Sedov-Taylor phase of the remnants is not well resolved. 
A number of other papers have reported closely related local simulations of SN feedback in vertically stratified media, with varying degrees of complexity  \citep[e.g.,][]{2006ApJ...638..797D, 2006ApJ...653.1266J, 2009ApJ...704..137J, 2012ApJ...754....2S, 2013MNRAS.429.1922C, 2014A&A...570A..81H, 2015ApJ...815...67K, 2016MNRAS.456.3432G}. 
Our study adds to this body of work by providing a  detailed analysis of the turbulence and wind properties as a function of gas surface density and SN seeding scheme, as well as an analysis of the limitations of local Cartesian box simulations for predicting the properties of galactic winds. 

Our main results can be summarised as follows:
\begin{itemize}
\item  
When the locations of SNe correlate with density peaks, the typical cooling radius of SNRs is smaller and hence individual SNe undergo stronger radiative losses.  
As a result, the corona dynamics are  sensitive to the SN seeding model, with the FX model in which more SNe explode in low-density regions above and below the mid-plane resulting in hotter galactic coronae and more powerful outflows. 
\item In all our simulations, hot ($T>10^{5}$ K) and/or low-density ($n_{\rm H}<0.01\bar{n}_{\rm H}$) gas makes up a negligible ($<10^{-3}$) mass fraction of the disc (Table \ref{tab:volfill}). 
Interestingly, the filling fraction of low density gas is generally larger than that of high temperature gas.
\item In agreement with simulations of isolated SNRs, the vast majority of the energy supplied by SNe is radiated away. The rate at which SNe energy goes into driving turbulence is $\dot{E}_{\rm turb} \lesssim 0.01\dot{E}_{\rm SN,tot}$, while outflows carry at most $\sim10\%$ of the total energy supplied by SNe (in the FX simulations; see Fig. \ref{fig:energetics_fx}). 
In the SC simulations, SNe couple less efficiently to the low-density material and  $\lesssim1\%$ of the total energy supplied by SNe goes into the wind (Fig. \ref{fig:energetics_sc}).
\item All of our simulations generate outflows from the computational domain.
However, the mass loading factors (ratio of outflow rate to star formation rate) $\eta \lesssim 1$ are low and the outflows remain subsonic throughout the computational domain.   
While our fiducial simulations are well converged with respect to spatial resolution (Appendix \ref{appendix:A}), the wind structure changes with increasing box height.  In particular, the outflow rate from the computational domain decreases significantly with increasing box height (see, e.g., Fig. \ref{fig:energetics_athena}).
These properties of the outflows in our simulations are contrary to the predictions of SNe-driven galactic winds \citep[e.g.,][]{CC85}.    We argue that this is primarily because there are in fact no adiabatic, steady state, subsonic to supersonic winds -- the standard SNe driven winds of galactic wind theory -- in Cartesian geometry (see \S \ref{sec:bc_dep} \& Appendix \ref{appendix:C}).   
\item Most of the outflowing material at a given height eventually falls back onto the disc: even at $\approx 0.5$ kpc above the mid-plane, only a very small fraction of the outflowing mass has sufficient velocity to potentially escape a realistic galaxy with escape velocity $\sim$300 km s$^{-1}$  ($\lesssim 1 \%$ even in the FX simulations; see Fig. \ref{fig:mass_loading_pdf}). These low mass loading factors are in tension with the larger wind mass loading  predicted by global galaxy simulations with more realistic geometry \citep[e.g.,][]{2012MNRAS.421.3522H, 2015MNRAS.454.2691M}; larger mass loading factors are also required in models that explain the observed galaxy stellar mass function and the metal enrichment of the intergalactic medium \citep[e.g.,][]{2006MNRAS.373.1265O, 2015ARA&A..53...51S}.   Because of the limitations of local simulations for studying galactic winds it is, however,  premature to conclude that this is a problem for SNe-driven galactic wind models.
\item The properties of the disc turbulence in our simulations are well converged with respect to both spatial resolution and box size. 
The mass-weighted turbulent velocity dispersion in the discs depends very weakly on disc properties. 
As in previous work (e.g., \citealt{2006ApJ...638..797D, 2009ApJ...704..137J, 2012ApJ...754....2S}), we find low mass-weighted turbulent velocities $\sigma_{\rm v,M}\approx3-7$ km s$^{-1}$ in all cases.  We explain this analytically in \S \ref{sec:turb} (see eq. \ref{eq:sigma_weak_function}). 
The weak dependence of the mass-weighted turbulent velocity dispersion on $\Sigma_{\rm g}$ is consistent with the quadratic Kennicutt-Schmidt law ($\dot{\Sigma}_{*} \propto \Sigma_{\rm g}^{2}$) predicted by models in which vertical gravity and stellar (SN) feedback roughly balance in galactic discs \citep[][]{2011ApJ...731...41O, 2013MNRAS.433.1970F}. 
\end{itemize}
Overall, we find that while local Cartesian box simulations are useful for studying how SNe excite turbulence and how/whether SNRs break out of galactic discs, they are less useful for predicting realistic galactic wind mass and energy loss rates. These limitations are likely generic to all local simulations in Cartesian geometry, not only ours.  In particular,  fitting functions for wind mass loading factors as a function of disc properties derived from  local calculations \citep[e.g.,][]{2013MNRAS.429.1922C} are unlikely to be robust. 

Our simulations also do not include the effects of galactic rotation and self-gravity. 
In real galaxies, rotation defines the \citet{1964ApJ...139.1217T} $Q$ parameter for gravitational stability of the disc. In many models of how star formation self-regulates \citep[e.g.,][]{1997ApJ...481..703S, 2005ApJ...630..167T, 2005ApJ...630..250K,2013MNRAS.433.1970F} it is assumed that discs reach a state of marginal stability, $Q\sim1$. 
Regulation to $Q\sim1$ is also supported by many global simulations of galactic discs (e.g., Dobbs et al. 2010\nocite{2010MNRAS.403..625D}; Hopkins et al. 2012\nocite{2012MNRAS.421.3488H}; Torrey et al., in prep.) and by some observations \citep[e.g.,][]{2011ApJ...733..101G}.
In a disc with $Q\sim1$, the turbulent velocity dispersion is directly related to the gas fraction and circular velocity $v_{\rm c}$ of the disc via $\sigma_{\rm v,M} \sim f_{\rm gas} v_{\rm c}$. 
For a massive gas-rich disc with $f_{\rm gas} \sim 0.5$ and $v_{\rm c}\sim250$ km s$^{-1}$, $Q \sim 1$ would thus predict a much higher turbulent velocity dispersion than the values $\sim 3-7$ km s$^{-1}$ we find in our  ULIRG-motivated simulations. 
Actual local ULIRGs and gas-rich $z\sim2$ star-forming galaxies are in fact observed to have elevated gas turbulent velocity dispersions $\sim50-100$ km s$^{-1}$ \citep[][]{1998ApJ...507..615D, 2011ApJ...733..101G} relative to more gas-poor local galaxies like the Milky Way.\footnote{More precisely, ULIRGs show larger non-thermal linewidths, even for molecular gas.   These are typically interpreted as arising from turbulent broadening but could also be due in part to non-circular gas kinematics.}   
This suggests that galactic rotation and self-gravity, not just stellar feedback, may be important ingredients in determining the turbulent velocity dispersions of real galactic discs.   It is not clear, however, whether these ingredients also influence how star formation self-regulates or whether the latter physics is relatively decoupled from the larger-scale rotational dynamics (see, e.g., \citealt{2012ApJ...754....2S}).

Finally, we note that including lower temperature cooling  and heating processes will change the phase structure of the ISM for some galaxy models, and potentially how SNe vent into the corona to drive a wind (see, e.g., \citealt{2015MNRAS.454..238W, 2016MNRAS.456.3432G}).    This may be important for quantitatively modeling galactic wind properties driven by SNe.   

To make further progress towards understanding the properties of turbulence and galactic winds driven by SNe, it will likely be necessary to more realistically model the global geometry and gravitational potential of galaxies, including perhaps the effects of disc rotation and self-gravity.   
Of course, in real galaxies other stellar feedback mechanisms (photoionization, stellar winds, radiation pressure, cosmic rays, ...) and additional physical processes such as magnetic fields and interstellar chemistry can also be important, and may be critical for understanding the science questions highlighted in this paper.   Isolating the effects of SNe alone is nonetheless important given their energetic importance for the ISM and galactic winds.  Our calculations  demonstrate that even the idealized problem of SN feedback alone still poses significant  challenges.

\section*{Acknowledgments}
We thank Todd Thompson and Paul Torrey for useful discussions.  DM was supported in part by the Swiss National Science Foundation postdoctoral fellowship and in part by NASA ATP grant 12-APT12-0183.  DF was supported by an NSF Graduate Research Fellowship.   CAFG was supported by NSF through grants AST-1412836 and AST-1517491, by NASA through grant NNX15AB22G, and by Northwestern University funds. 
EQ was supported in part by NASA ATP grant 12-APT12-0183, a Simons Investigator award from the Simons Foundation, and the David and Lucile Packard Foundation. 
The simulations reported in this paper were run and processed on the Savio computer cluster at UC Berkeley and with resources provided by the NASA High-End Computing (HEC) Program through the NASA Advanced Supercomputing (NAS) Division at Ames Research Center (allocations SMD-14-5492, SMD-14-5189, and SMD-15-6530).


\bibliography{main}


\appendix
\section{Resolution test}\label{appendix:A}

\begin{figure}
    \includegraphics[width=0.49\textwidth]{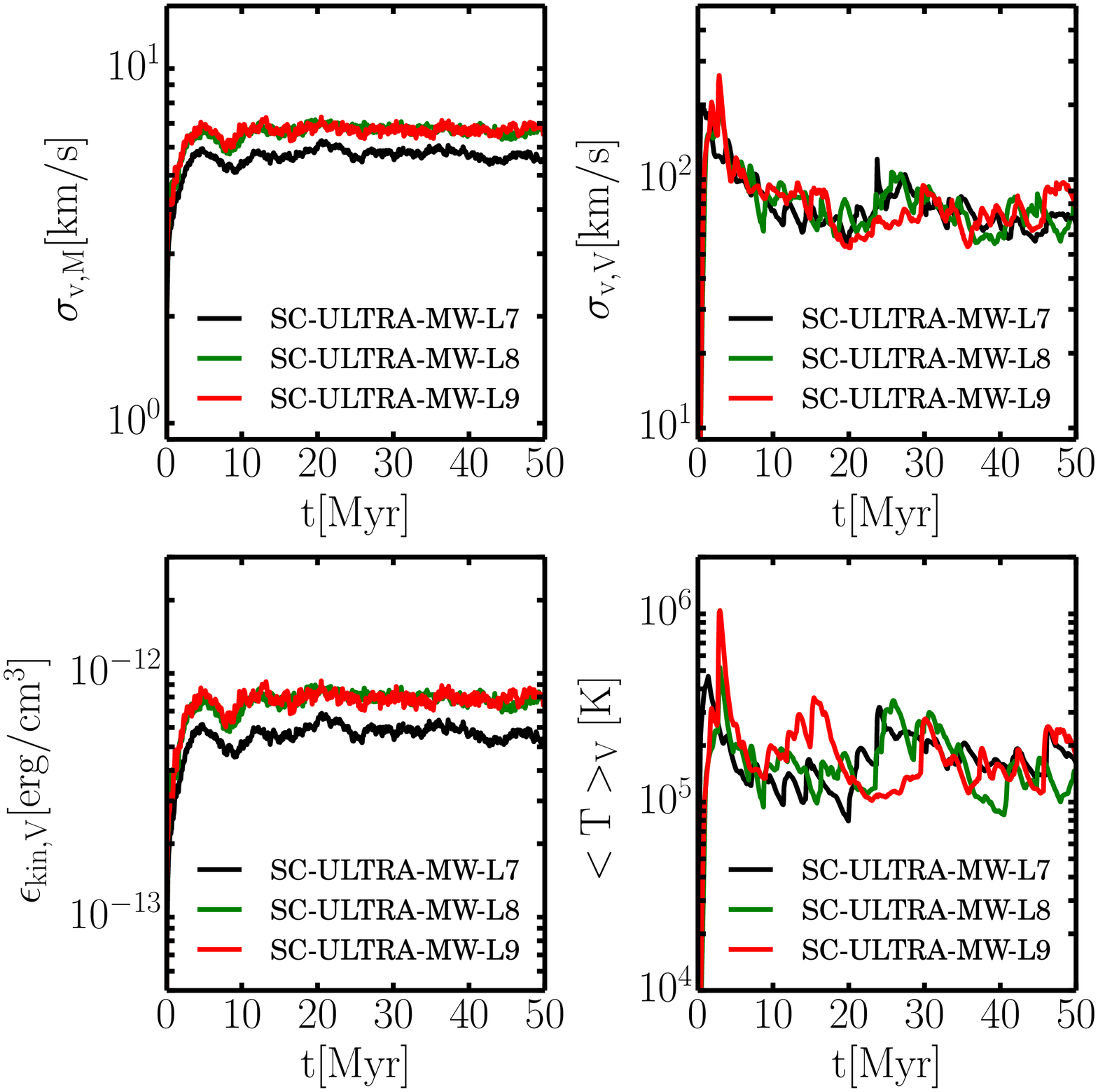}
\caption{ Resolution test:  comparison of several bulk properties of the simulations (averaged over the whole domain) as a function of time for SC-ULTRA-MW-L7 (black), SC-ULTRA-MW-L8 (green) and SC-ULTRA-MW-L9 (red).  Top left: mass weighted velocity dispersion. Top right: volume weighted velocity dispersion. Bottom left: kinetic energy density. Bottom right: volume weighted temperature.   By our L8 simulation which has a cell size $\Delta x = 3.9$ pc, the results appear well converged.}\label{fig:integrated_resolution}
\end{figure}

\begin{figure}
    \includegraphics[width=0.49\textwidth]{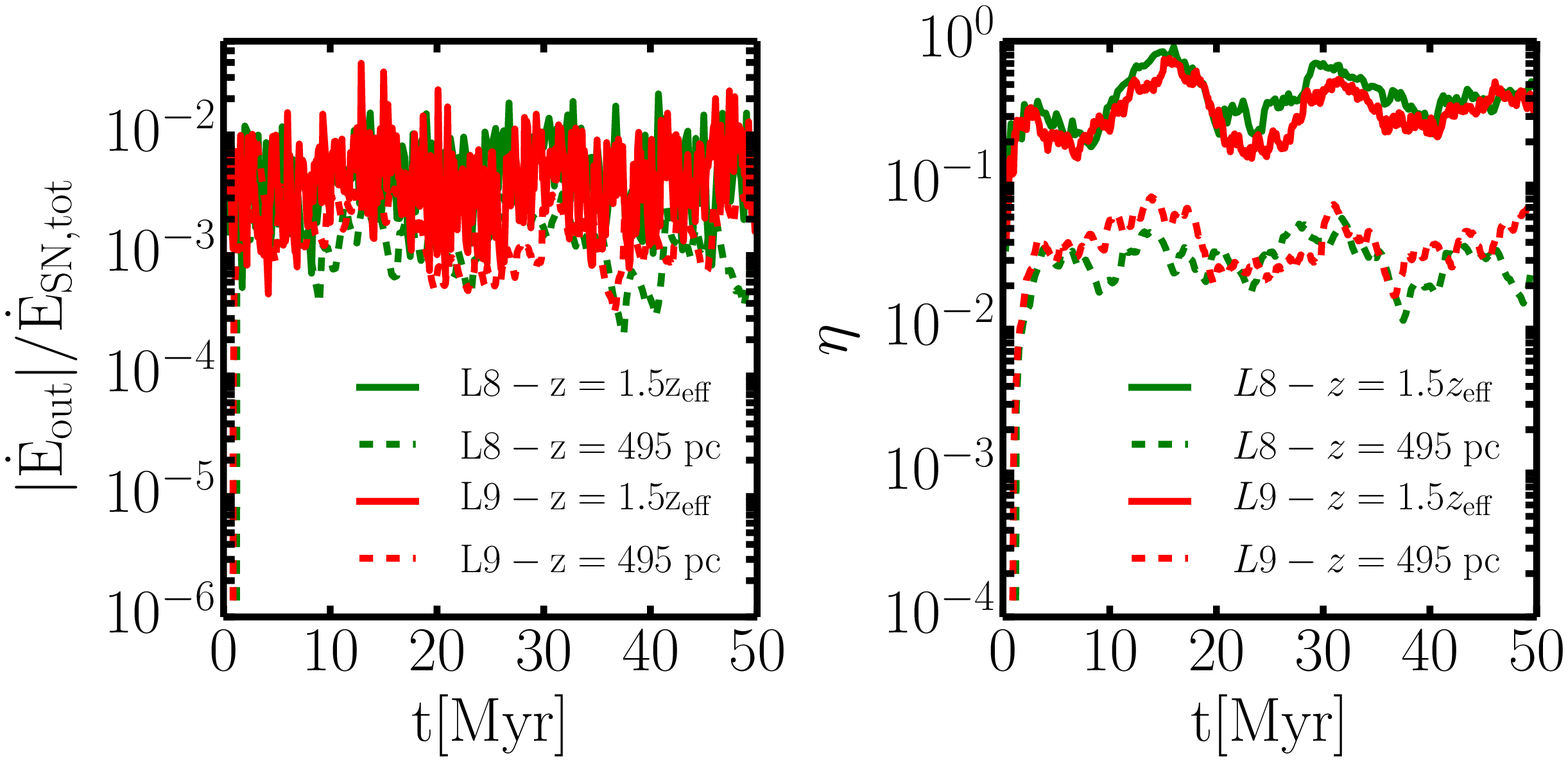}
    \caption{Dependence of the wind energy outflow rate (left) and mass loading factor $\eta$ (right) on resolution, for the SC-ULTRA-MW-L8 (green) and SC-ULTRA-MW-L9 (red) simulations.  These global properties of the outflow are reasonably well converged. }
\label{fig:energetics_resolution}
\end{figure}

Figures~\ref{fig:integrated_resolution} \& \ref{fig:energetics_resolution} compare 3 versions of the ULTRA-MW simulations with SC SN seeding: SC-ULTRA-MW-L7, SC-ULTRA-MW-L8 and SC-ULTRA-MW-L9 with spatial resolution (cell size) 7.8 pc, 3.9 pc, and 1.95 pc, respectively. The simulations show good 
convergence, both in the properties of the turbulence and in the outflow properties.   Figure~\ref{fig:integrated_resolution} shows that a resolution of at least 3.9 pc (-L8) is preferred for convergence in some of the volume weighted quantities.

\section{Dependence of Outflow Properties on Vertical Height}\label{appendix:B}

\begin{figure}
    \includegraphics[width=0.49\textwidth]{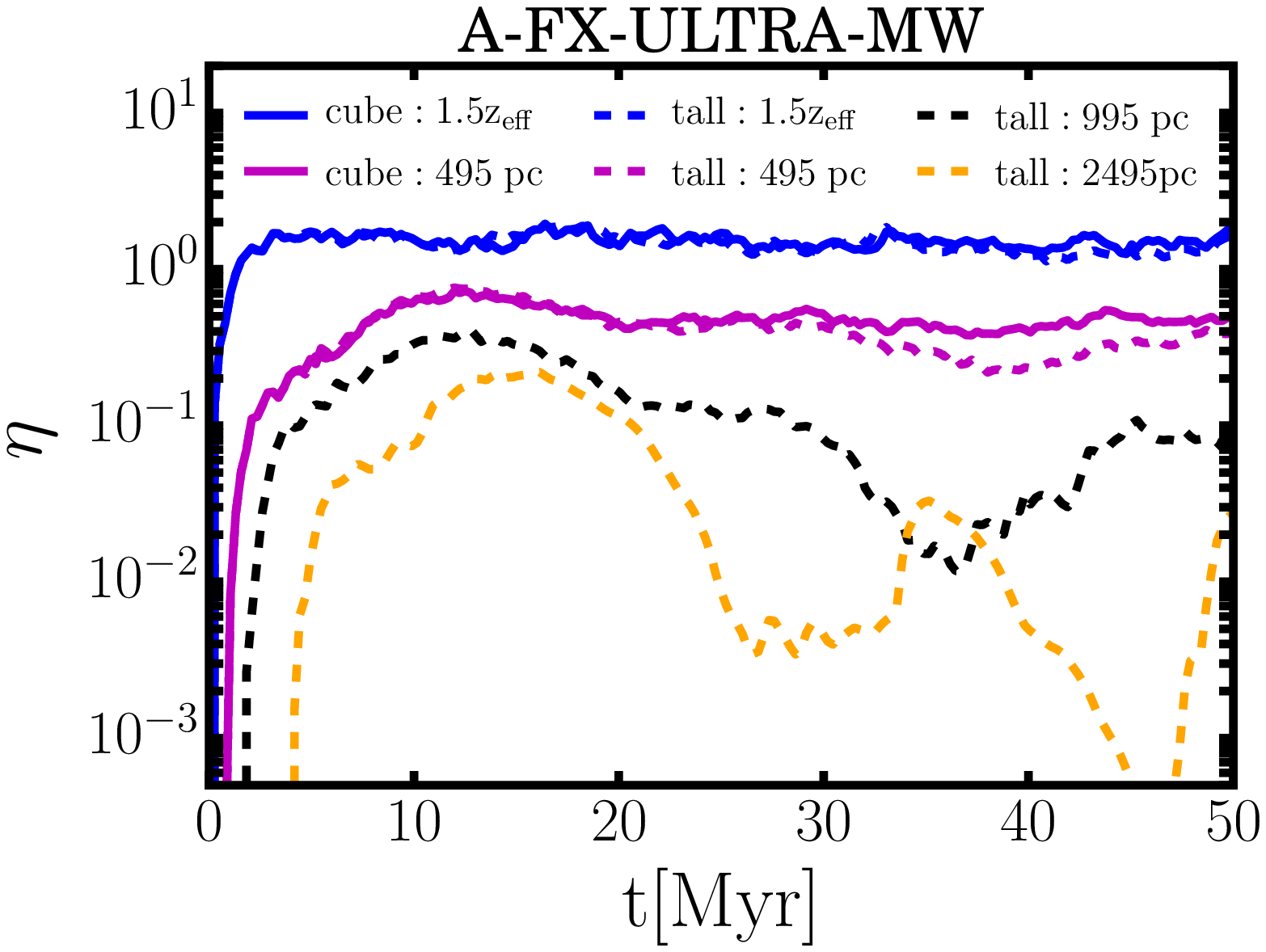}
\caption{Effect of box size on the wind mass loading factor $\eta$ (the ratio of outflow rate to star formation rate) at different heights, for the ULTRA-MW model (as in Fig. \ref{fig:energetics_lb} of the main text).   These simulations are with \textsc{athena}.  We compare a cubic box like that used in the main text (1 kpc$^3$) to a tall box (1 kpc $\times$ 1 kpc $\times$ 5 kpc).  The outflow rate is substantially lower at the boundary of the tall box, confirming the conclusion of \S \ref{sec:bc_dep} that there is not a well-defined (i.e., independent of box properties) outflow rate in these local Cartesian simulations.}\label{fig:energetics_athena}
\end{figure}

The comparison of our fiducial  simulations to the large box simulations in \S \ref{sec:bc_dep} indicates that the location of the vertical boundary  impacts the outflows' properties.  To test this in more detail it is useful to put this vertical boundary  even farther from the disc midplane. Unfortunately, \textsc{ramses} requires that the computational domain be cubical, so to extend the vertical boundary we must also extend the horizontal boundaries, which increases the computational cost and is not necessarily equivalent to simply increasing the vertical size of the domain. To avoid this issue we implemented the exact same SN feedback in a stratified medium setup in  \textsc{athena} \citep{2008ApJS..178..137S}, which has very similar capabilities, but for this problem has the benefit of being able to use arbitrary rectangular domains. 

We ran two simulations with \textsc{athena}. In the first we exactly duplicate FX-ULTRA-MW-L8. It has a $256^3$ cubical domain, 1 kpc on a side, and we refer to it as A-FX-ULTRA-MW-L8. The second has the same fixed SN injection scheme, and the same initial conditions, but is elongated along the $z$-axis. This run, which we refer to as A-FX-ULTRA-MW-TALL, is 1 kpc $\times$ 1 kpc $\times$ 5 kpc in size and has 256 $\times$ 256 $\times$ 1,280 cells. Figure \ref{fig:energetics_athena} shows the mass loading of the outflow for both of these simulations. 

Reassuringly, the differences between the matching \textsc{ramses} and \textsc{athena} simulations, A-FX-ULTRA-MW-L8 and FX-ULTRA-MW-L8, are negligible. The minor differences arise due to slight differences in the numerical techniques used in the two codes and stochastic differences in SN seeding.    The key point of Figure \ref{fig:energetics_athena} is that the outflow rate continues to decrease significantly at large $z$, confirming our conclusion in \S \ref{sec:bc_dep} that there is not a well defined outflow rate for the local Cartesian simulations.   

\section{Analytics of winds in stratified geometries}\label{appendix:C}
The stratified Cartesian boxes utilized in this and previous studies allow one to study the dynamics of overlapping SNe at  higher resolution than is possible with comparable computational time in fully global models of galactic discs.  One downside, however, is that the absence of a realistic global geometry precludes standard steady subsonic to supersonic winds from developing (related concerns have been noted in the shearing box accretion disc literature; e.g., \citealt{Fromang2013}).

To see this we consider a steady state wind in the spirit of \cite{CC85}, in which SNe produce a mass per unit time per unit volume of $q$ and an energy per unit mass of $\epsilon$.  For a steady flow in the $z$ direction in Cartesian geometry, the equations of conservation of mass, momentum, and entropy become
\begin{equation}
\frac{d}{dz}\left(\rho v \right) = q,
\label{eq:App-mass}
\end{equation}
\begin{equation}
\rho v \frac{dv}{dz} = -\frac{dP}{dz} - \rho \frac{d \Phi}{dz} - qv = 0,
\label{eq:App-mom}
\end{equation}
and
\begin{equation}
\rho v T \frac{ds}{dz} = q\left(\frac{1}{2} v^2 + \epsilon -\frac{5}{2}\frac{P}{\rho}\right) - \rho^2 \tilde \Lambda,
\label{eq:App-entropy}
\end{equation}
where $v$ is the vertical flow speed, we have assumed an ideal gas with an adiabatic index of $\gamma = 5/3$, and $\tilde \Lambda = \Lambda/(\mu m_p)^2$.   

Equations \ref{eq:App-mass}-\ref{eq:App-entropy} can be combined using standard manipulations to yield a wind equation for the vertical velocity:
\begin{equation}
\frac{\rho}{v}\frac{dv}{dz}\left(v^2 - \frac{5}{3}\frac{P}{\rho}\right) = - \rho \frac{d\Phi}{dz} - \frac{4}{3} q v - \frac{2}{3} \frac{q \epsilon}{v} + \frac{2}{3} \frac{\rho^2 \tilde \Lambda}{v}
\label{eq:App-wind}
\end{equation}
A nominal sonic point is associated with $v^2 = 5/3 (P/\rho)$ and so the right hand side of Equation~\ref{eq:App-wind} must vanish at the sonic point.  However, all of the terms on the right hand side of Equation~\ref{eq:App-wind} are negative except the cooling term $\propto \tilde \Lambda$.  This implies that in Cartesian geometry there can only be a sonic point in the presence of strong cooling.  Adiabatic steady state winds which undergo a subsonic to supersonic transition -- the standard supernovae driven winds of galactic wind theory \citep{CC85} -- do not exist in Cartesian geometry.  This is simply because of the lack of the $1/r^2$ spherical divergence term that usually appears on the right hand side of Equation~\ref{eq:App-wind}.  As discussed in \S \ref{sec:bc_dep}, the absence of such wind solutions in  Cartesian geometry calls into question {quantitative} results on galactic winds derived from stratified Cartesian simulations.


\label{lastpage}
\end{document}